\documentclass[11pt,letter]{article}

\usepackage[letterpaper,hmargin=0.97in,vmargin=1in]{geometry}

\usepackage{times}

\usepackage{amssymb,amsmath,stmaryrd}
\usepackage[lined,boxed,commentsnumbered]{algorithm2e}
\newtheorem{theorem}{Theorem}[section]
\newtheorem{dfn}[theorem]{Definition}
\newtheorem{lemma}[theorem]{Lemma}
\newtheorem{claim}[theorem]{Claim}

\newcommand{\sq}{\hbox{\rlap{$\sqcap$}$\sqcup$}}
\newcommand{\qed}{\hspace*{\fill}\sq}
\newenvironment{proof}{\noindent {\bf Proof.}\ }{\qed\par\vskip 4mm\par}

\def\reals{{\mathbb{R}}}

\def\BR{{\mathcal{BR}}}

\newcommand{\vo}{\mathbf{0}}

\title{Approximate Pure Nash Equilibria in Weighted Congestion Games: Existence, Efficient Computation, and Structure\thanks{This work was partially supported by the grant NRF-RF2009-08 ``Algorithmic aspects of coalitional games'' and the EC-funded STREP Project FP7-ICT-258307 EULER.}}

\author{Ioannis Caragiannis\thanks{Computer Technology Institute \& Department of Computer Engineering and Informatics, University of Patras, Greece. Email: {\tt caragian@ceid.upatras.gr}} \and Angelo Fanelli\thanks{Division of Mathematical Sciences, School of Physical and Mathematical Sciences, Nanyang Technological University, Singapore. Email: {\tt angelo.fanelli@ntu.edu.sg, ngravin@pmail.ntu.edu.sg}}\\ \and Nick Gravin$^\ddag$\\ \and Alexander Skopalik\thanks{TU Dortmund University, Germany. Email: {\tt Alexander.Skopalik@tu-dortmund.de}}}

\begin{document}

\maketitle

\begin{abstract}
We consider structural and algorithmic questions related to the Nash dynamics of weighted congestion games. In weighted congestion games with linear latency functions, the existence of pure Nash equilibria is guaranteed by potential function arguments. Unfortunately, this proof of existence is inefficient and computing pure Nash equilibria is such games is a {\sf PLS}-hard problem even when all players have unit weights. The situation gets worse when superlinear (e.g., quadratic) latency functions come into play; in this case, the Nash dynamics of the game may contain cycles and pure Nash equilibria may not even exist. Given these obstacles, we consider approximate pure Nash equilibria as alternative solution concepts. Do such equilibria exist? And if so, can we compute them efficiently?

We provide positive answers to both questions for weighted congestion games with polynomial latency functions by exploiting an ``approximation'' of such games by a new class of potential games that we call $\Psi$-games. This allows us to show that these games have $d!$-approximate pure Nash equilibria, where $d$ is the maximum degree of the latency functions. Our main technical contribution is an efficient algorithm for computing $O(1)$-approximate pure Nash equilibria when $d$ is a constant. For games with linear latency functions, the approximation guarantee is $\frac{3+\sqrt{5}}{2}+O(\gamma)$ for arbitrarily small $\gamma>0$; for latency functions with maximum degree $d\geq 2$, it is $d^{2d+o(d)}$. The running time is polynomial in the number of bits in the representation of the game and $1/\gamma$. As a byproduct of our techniques, we also show the following interesting structural statement for weighted congestion games with polynomial latency functions of maximum degree $d\geq 2$: polynomially-long sequences of best-response moves from any initial state to a $d^{O(d^2)}$-approximate pure Nash equilibrium exist and can be efficiently identified in such games as long as $d$ is constant.

To the best of our knowledge, these are the first positive algorithmic results for approximate pure Nash equilibria in weighted congestion games. Our techniques significantly extend our recent work on unweighted congestion games through the use of $\Psi$-games. The concept of approximating non-potential games by potential ones seems to be novel and might have further applications.
\end{abstract}

\newpage
\section{Introduction}
\label{sec:intro}
Among other solution concepts, the notion of the pure Nash equilibrium plays a central role in Game Theory. Pure Nash equilibria in a game characterize situations with non-cooperative deterministic players in which no player has any incentive to unilaterally deviate from the current situation in order to achieve a higher payoff. Unfortunately, it is well known that there are games that do not have pure Nash equilibria. Furhermore, even in games where the existence of equilibria is guaranteed, their computation can be a computationally hard task. Such negative results significantly question the importance of pure Nash equilibria as solution concepts that characterize the behavior of rational players.

Approximate pure Nash equilibria, which characterize situations where no player can {\em significantly improve} her payoff by unilaterally deviating from her current strategy, could serve as alternative solution concepts\footnote{Actually, approximate pure Nash equilibria may be more desirable as solution concepts in practical decision making settings since they can accommodate small modeling inaccuracies due to uncertainty (e.g., see the arguments in \cite{CGC04}).} provided that they exist and can be computed efficiently. In this paper, we present the first positive algorithmic results for approximate pure Nash equilibria in weighted congestion games. Our main contribution is a polynomial-time algorithm that computes $O(1)$-approximate pure Nash equilibria under mild restrictions on the game parameters; these restrictions apply to important subclasses of games in which not even the existence of such approximate equilibria was known prior to our work.

\medskip\noindent{\bf Problem statement and related work.}
In a weighted congestion game, players compete over a set of resources. Each player has a positive weight. Each resource incurs a latency to all players that use it; this latency depends on the total weight of the players that use the resource according to a resource-specific, non-negative, and non-decreasing latency function. Among a given set of strategies (over sets of resources), each player aims to select one selfishly, trying to minimize her individual total cost, i.e., the sum of the latencies on the resources in her strategy. Typical examples include weighted congestion games in networks, where the network links correspond to the resources and each player has alternative paths that connect two nodes as strategies.

The case of unweighted congestion games (i.e., when all players have unit weight) has been widely studied in literature. Rosenthal \cite{R73} proved that these games admit a potential function with the following remarkable property: the difference in the potential value between two states (i.e., two snapshots of strategies) that differ in the strategy of a single player equals to the difference of the cost experienced by this player in these two states. This immediately implies the existence of a pure Nash equilibrium. Any sequence of improvement moves by the players strictly decreases the value of the potential and a state corresponding to a local minimum of the potential will eventually be reached; this corresponds to a pure Nash equilibrium. For weighted congestion games, potential functions exist only in the case where the latency functions are linear or exponential (see \cite{FKS05,HK10,PS06}). Actually, in games with polynomial latency functions (of constant maximum degree higher than $1$), pure Nash equilibria may not exist \cite{HK10}. In general, the problem of deciding whether a given weighted congestion game has a pure Nash equilibrium is {\sf NP}-hard \cite{DS08}.

Potential functions provide only inefficient proofs of existence of pure Nash equilibria. Fabrikant et al. \cite{FabrikantPT04} proved that the problem of computing a pure Nash equilibrium in a (unweighted) congestion game is {\sf PLS}-complete (informally, as hard as it could be given that there is an associated potential function; see \cite{Johnson88}). This negative result holds even in the case of linear latency functions \cite{AckermannRV08}. One consequence of {\sf PLS}-completeness results is that almost all states in some congestion games are such that any sequence of players' improvement moves that originates from these states and reaches pure Nash equilibria is exponentially long. Such phenomena have been observed even in very simple weighted congestion games (see \cite{ARV09,EKM07}). Efficient algorithms are known only for special cases. For example, Fabrikant et al. \cite{FabrikantPT04} show that the Rosenthal's potential function can be (globally) minimized efficiently by a flow computation in unweighted congestion games in networks when the strategy sets of the players are symmetric.

The above negative results have led to the study of the complexity of approximate pure Nash equilibria (or, simply, approximate equilibria). A $\rho$-approximate (pure Nash) equilibrium is a state, from which no player has an incentive to deviate so that she decreases her cost by a factor larger than $\rho$. In our recent work \cite{CFGS11}, we present an algorithm for computing $O(1)$-approximate equilibria for unweighted congestion games with polynomial latency functions of constant maximum degree. The restriction on the latency functions is necessary since, for more general latency functions, Skopalik and V\"ocking \cite{SkopalikV08} show that the problem is still {\sf PLS}-complete for any polynomially computable $\rho$ (see also the discussion in \cite{CFGS11}). Improved bounds are known for special cases. For symmetric unweighted congestion games, Chien and Sinclair \cite{ChienS07} prove that the $(1+\epsilon)$-improvement dynamics converges to a $(1 + \epsilon$)-approximate equilibrium after a polynomial number of steps; this result holds under mild assumptions on the latency functions and the participation of the players in the dynamics. Efficient algorithms for approximate equilibria have been recently obtained for other classes of games such as constraint satisfaction \cite{BhalgatCK10,NT09}, anonymous games \cite{DP07}, network formation \cite{AC09}, and facility location games \cite{CH10}.

In light of the negative results mentioned above, several authors have considered other properties of the dynamics of congestion games. The papers \cite{AwerbuchAEMS08,FM09, GMV05} consider the question of whether efficient states (in the sense that the total cost of the players, or social cost, is small compared to the optimum one) can be reached by best-response moves in linear weighted congestion games. In particular, Awerbuch et al. \cite{AwerbuchAEMS08} show that using almost unrestricted sequences of ($1+\epsilon$)-improvement best-response moves, the players rapidly converge to efficient states. Unfortunately, these states are not approximate equilibria, in general. Similar approaches have been followed in the context of other games as well, such as multicast \cite{CKM+08,CCL+07}, cut \cite{CMS06}, and valid-utility games \cite{MV04}.

\medskip\noindent{\bf Our contribution.} To the best of our knowledge, no efficient algorithm for computing approximate equilibria is known for (any broad enough subclass of) weighted congestion games. We fill this gap by presenting an algorithm for computing $O(1)$-approximate equilibria in weighted congestion games with polynomial latency functions of constant maximum degree. For games with linear latency functions, the approximation guarantee is $\frac{3+\sqrt{5}}{2}+O(\gamma)$ for arbitrarily small $\gamma>0$; for latency functions of maximum degree $d\geq 2$, it is $d^{2d+o(d)}$. The algorithm runs in time that is polynomial in the number of bits in the representation of the game and $1/\gamma$.

This result is much more surprising than it looks at first glance. In particular, weighted congestion games with superlinear latency functions do not admit potential functions, the main tool that is exploited by all known positive algorithmic results for (approximate) equilibria in congestion games. Given this, it is not even clear that $O(1)$-approximate equilibria exist. In order to bypass this obstacle, we introduce a new class of potential games (that we call $\Psi$-games), which ``approximate'' weighted congestion games with polynomial latency functions in the following sense. $\Psi$-games of degree $1$ are linear weighted congestion games. Each weighted congestion game of degree $d\geq 2$ has a corresponding $\Psi$-game of degree $d$ defined in such a way that any $\rho$-approximate equilibrium in the latter is a $d!\rho$-approximate equilibrium for the former. As an intermediate new result, we obtain that weighted congestion games with polynomial latency functions of degree $d$ have $d!$-approximate equilibria.

So, our algorithm is actually applied to $\Psi$-games. It has a simple general structure, similar to our recent algorithm for unweighted congestion games \cite{CFGS11}, but has also important differences that are due to the dependency of the cost of each player on the weights of other players. Given a $\Psi$-game of degree $d$ and an arbitrary initial state, the algorithm computes a sequence of best-response player moves of length that is bounded by a polynomial in the number of bits in the representation of the game and $1/\gamma$. The sequence consists of phases so that the players that participate in each phase experience costs that are polynomially related. This is crucial in order to obtain convergence in polynomial time. Within each phase, the algorithm coordinates the best-response moves according to two different but simple criteria; this is the main tool that guarantees that the effect of a phase to previous ones is negligible and, eventually, an approximate equilibrium is reached. The approximation guarantee is slightly higher than a quantity that characterizes the potential functions of $\Psi$-games; this quantity (which we call the {\em stretch}) is defined as the worst-case ratio of the potential value at an almost exact pure Nash equilibrium over the globally optimum potential value and is almost $\frac{3+\sqrt{5}}{2}$ for linear weighted congestion games and $d^{d+o(d)}$ for $\Psi$-games of degree $d\geq 2$. Our analysis follows the same main steps as in our recent paper \cite{CFGS11} but uses significantly more involved arguments due to the definition of $\Psi$-games.

We also present a similar but slightly inferior algorithm that is applied directly to weighted congestion games of maximum degree $d\geq 2$ and reveals a rather surprising structural property of their Nash dynamics: starting from any initial state, the algorithm identifies a polynomially-long sequence of best-response moves that lead to a $d^{O(d^2)}$-approximate equilibrium. Even though the definition of this algorithm does not make any use of properties of $\Psi$-games, the analysis is heavily based on them, similarly to the analysis of our main algorithm.

We remark that, following the classical definition of polynomial latency functions in the literature, we assume that they have non-negative coefficients. This is a necessary limitation since the problem of computing a $\rho$-approximate equilibrium in (unweighted) congestion games with linear latency functions with negative offsets is {\sf PLS}-complete for any polynomial-time computable $\rho\geq 1$ \cite{CFGS11}.

\medskip\noindent{\bf Roadmap.} We begin with preliminary general definitions in Section \ref{sec:prelim}. Section \ref{sec:psi-games} is devoted to $\Psi$-games and their properties. We present our algorithm and its analysis in Section \ref{sec:algo} and conclude with open problems in Section \ref{sec:open}. Due to lack of space, most of the proofs as well as our structural result appears in Appendix.

\section{Definitions and preliminaries}\label{sec:prelim}
In general, a \emph{game} can be defined as follows. It has a set of $n$ players ${\cal N}$; each player $u\in {\cal N}$ has a set of available strategies $\Sigma_u$. A snapshot of strategies, with one strategy per player, is called a {\em state}. Each state $S\in \prod_{u\in {\cal N}}{\Sigma_u}$ incurs a positive cost $c_u(S)$ to player $u$. Players act selfishly; each of them aims to select a strategy that minimizes her cost, given the strategies of the other players. Given a state $S$ and a strategy $s'_u\in \Sigma_u$ for player $u$, we denote by $(S_{-u},s'_u)$ the state obtained from $S$ when player $u$ {\em deviates} to strategy $s'_u$. For a state $S$, an {\em improvement move} (or, simply, a {\em move}) for player $u$ is the deviation to any strategy $s'_u$ that (strictly) decreases her cost, i.e., $c_u(S_{-u},s'_u) < c_u(S)$. For $\rho\geq 1$, such a move is called a $\rho$-{\em move} if it satisfies $c_u(S_{-u},s'_u) < \frac{c_u(S)}{\rho}$. A {\em best-response move} is a move that minimizes the cost of the player (of course, given the strategies of the other players). So, from state $S$, a move of player $u$ to strategy $s_u$ is a best-response move (and is denoted by $\BR_u(S)$) when $c_u(S_{-u},s'_u) = \min_{s\in \Sigma_u}c_u(S_{-u},s)$. A state $S$ is called a {\em pure Nash equilibrium} (or, simply, an {\em equilibrium}) when $c_u(S)\leq c_u(S_{-u},s'_u)$ for every player $u\in {\cal N}$ and every strategy $s'_u\in \Sigma_u$, i.e., when no player has a move. In this case, we say that no player has (any incentive to make) a move. Similarly, a state is called a $\rho$-{\em approximate pure Nash equilibrium} (henceforth called, simply, a $\rho$-{\em approximate equilibrium}) when no player has a $\rho$-move. Also, a state is called a $\rho$-approximate equilibrium for a subset of players $A\subseteq {\cal N}$ if no player in $A$ has a $\rho$-move. We use the term {\em Nash dynamics} of a game in order to refer to the directed graph with nodes that correspond to the possible states of the game and directed edges that indicate improvement player moves; pure Nash equilibria correspond to sinks of the Nash dynamics.

A \emph{weighted congestion game} ${\cal G}$ can be represented by the tuple $\left(N, E, (w_u)_{u\in {\cal N}}, (\Sigma_u)_{u \in {\cal N}}, (f_e)_{e \in E}\right)$. There is a set of $n$ {\em players} ${\cal N}$ and a set of {\em resources} $E$. Each player $u$ has a positive weight $w_u$ and a set of available {\em strategies} $\Sigma_u$; each strategy $s_u$ in $\Sigma_u$ consists of a non-empty set of resources, i.e., $s_u\subseteq 2^E$. Each resource $e\in E$ has a non-negative and non-decreasing {\em latency function} $f_e$ defined over non-negative reals, which denotes the latency incurred to the players using resource $e$; this latency depends on the total weight of players whose strategies include the particular resource. For a state $S$, let us define $N_e(S)$ to be the multi-set of the weights of the players that use resource $e$ in $S$, i.e., $N_e(S)=\{w_u: u\in {\cal N} \mbox{ such that } e\in s_u\}$. Also, we use the notation $L(A)$ to denote the sum of the elements of a finite multi-set of reals $A$. Then, the latency incurred by resource $e$ to a player $u$ that uses it is $f_e(L(N_e(S)))$. The {\em cost} of a player $u$ at a state $S$ is the total latency she experiences at the resources in her strategy $s_u$ multiplied by her weight, i.e., $c_u(S)=w_u\sum_{e\in s_u}{f_e(L(N_e(S)))}$. We consider weighted congestion games in which the resources have polynomial latency functions with (integer) maximum degree $d\geq 1$ with non-negative coefficients. More precisely, the latency function of resource $e$ is $f_e(x) = \sum_{k=0}^d{a_{e,k}x^k}$ with $a_{e,k}\geq 0$. The special case of linear weighted congestion games (i.e., with latency functions of degree $1$) is of particular interest.
In general, the size of the representation of a weighted congestion game is the number of bits required to represent the parameters $a_{e,k}$ of the latency functions, the weights of the players, and their strategy sets. In weighted congestion games in networks, the network links are the resources. Each player $u$ aims to connect a pair of nodes $(s_u,t_u)$ and her strategies are all paths connecting $s_u$ with $t_u$ in the network. Note that the representation of such games does not need to keep the whole set of strategies explicitly; it just has to represent the parameters $a_{e,k}$, the weight and the source-destination node pair of each player, and the network.

Unweighted congestion games (i.e., when $w_u=1$ for each player $u\in{\cal N}$) as well as linear weighted congestion games are potential games. They admit a {\em potential function} $\Phi:\prod_u{\Sigma_u}\mapsto \reals^+$, defined over all states of the game, with the following property: for any two states $S$ and $(S_{-u},s'_u)$ that differ only in the strategy of player $u$, it holds that $\Phi(S_{-u},s'_u) - \Phi(S) = c_u(S_{-u},s'_u)-c_u(S)$. Clearly, the local minima of the potential function correspond to states that are pure Nash equilibria. The existence of a potential function also implies that the Nash dynamics of the corresponding game is acyclic. Potential functions for the two classes of games mentioned above have been presented by Rosenthal \cite{R73} and Fotakis et al. \cite{FKS05}, respectively. Unfortunately, weighted congestion games with polynomial latency functions of degree at least $2$ are not potential games and may not even have pure Nash equilibria \cite{HK10}.

\section{$\Psi$-games}\label{sec:psi-games}
Our aim in this section is to define a new class of games which we call $\Psi$-games and study their properties. We will need the following interesting family of functions which have also been used in \cite{C09} in a slightly different context.

\begin{dfn}
For integer $k\geq 0$, the function $\Psi_k$ mapping finite multi-sets of reals to the reals is defined as follows:
$\Psi_k(\emptyset)=0$ for any integer $k\geq 1$, $\Psi_0(A)=1$ for any (possibly empty) multi-set $A$, and for any
non-empty multi-set $A=\{\alpha_1, \alpha_2, ..., \alpha_\ell\}$ and integer $k\geq 1$,
\[\Psi_k(A)=k! \sum_{1\leq d_1 \leq ... \leq d_k \leq \ell}{\prod_{t=1}^{k}{\alpha_{d_t}}}.\]
\end{dfn}
So, $\Psi_k(A)$ is essentially the sum of all monomials of total degree $k$ on the elements of $A$. Each
term in the sum has coefficient $k!$. Clearly, $\Psi_1(A)=L(A)$. For $k\geq 2$, compare $\Psi_k(A)$ with $L(A)^k$
which can also be expressed as the sum of the same terms, albeit with different coefficients in $\{1, ..., k!\}$,
given by the multinomial theorem.

We are ready to define $\Psi$-games. A $\Psi$-game ${\cal G}$ of (integer) degree $d\geq 1$ can be represented by the tuple $({\cal N},E,(w_u)_{u\in {\cal N}},(\Sigma_u)_{u\in {\cal N}},(a_{e,k})_{e\in E, k=0, 1, ..., d})$. Similarly to weighted congestion games, there is a set of $n$ players ${\cal N}$ and a set of resources $E$. Each player $u$ has a weight $w_u$ and a set of available strategies $\Sigma_u$; each strategy $s_u \in\Sigma_u$ consists of a non-empty set of resources, i.e., $s_u\subseteq 2^E$. Each resource $e$ is associated with $d+1$ non-negative numbers $a_{e,k}$ for $k=0, 1, ..., d$. Again, for a state $S$, we define $N_e(S)$ to be the multi-set of weights of the players that use resource $e$ at state $S$. Then, the cost of a player $u$ at a state $S$ is defined as
$$\hat{c}_u(S) = w_u\sum_{e\in s_u}{\sum_{k=0}^d{a_{e,k}\Psi_k(N_e(S))}}.$$
Of course, the general definitions in the beginning of Section \ref{sec:prelim} apply also to $\Psi$-games. With some abuse in notation, we also use $\vo$ to refer to the pseudo-state in which no player selects any strategy and $\BR_u(\vo)$ to denote the best-response of player $u$ assuming that no other player participates in the game.

Clearly, given a weighted congestion game with polynomial latency functions of maximum degree $d$, there is a corresponding $\Psi$-game with degree $d$, i.e., the one with the same sets of players, resources, and strategy sets, and parameter $a_{e,k}$ for each resource $e$ and integer $k=0, 1, ..., d$ equal to the corresponding coefficient of the latency function $f_e$. Observe that $\Psi$-games of degree $1$ are linear weighted congestion games. As we will see below, in a sense, a $\Psi$-game of degree $d\geq 2$ is an approximation of its corresponding weighted congestion game.

We remark here that a different approximation of weighted congestion games has been recently considered by Kollias and Roughgarden \cite{KR11}. Given a weighted congestion game, they define a new game by answering the following question: how should the product of the total weight of the players that use the resource times its latency be shared as cost among these players so that the resulting game is a potential game? Their games use a different sharing than the weight-proportional one used by weighted congestion games. In contrast, our approach is to define an artificial latency on each resource $e$ (by replacing the term $a_{e,k}L(N_e(S))^k$ with $a_{e,k}\Psi_k(N_e(S))$ in the latency functions) so that weight-proportional sharing yields a potential game. This guarantees the relation between approximate equilibria in weighted congestion games and $\Psi$-games presented in Lemma \ref{lem:approx} below, which is crucial for our purposes.

\medskip\noindent{\bf Properties of $\Psi$-games.} We begin with a very important property of $\Psi$-games.
\begin{theorem}\label{thm:psi-potential}
The function $\Phi(S) = \sum_e{\sum_{k=0}^d{\frac{a_{e,k}}{k+1}\Psi_{k+1}(N_e(S))}}$ is a potential function for $\Psi$-games of degree $d$.
\end{theorem}

As a corollary, we conclude that the Nash dynamics of $\Psi$-games are acyclic; hence, these games admit pure Nash equilibria. Recall that $\Psi$-games of degree $1$ are linear weighted congestion games; for this specific case, Theorem \ref{thm:psi-potential} has been proved in \cite{FKS05}.

In the following, we study the relation between the approximation guarantee of a state for a $\Psi$-game and its corresponding weighted congestion game with polynomial latency functions. 
%The proof of the next claim follows easily by Lemma \ref{lem:properties}a.
\begin{claim}\label{claim:approx}
Consider a weighted congestion game with polynomial latency functions of degree $d$ and its corresponding $\Psi$-game. Then, for each player $u$ and state $S$, $c_u(S)\leq \hat{c}_u(S)\leq d! c_u(S)$.
\end{claim}

Using Claim \ref{claim:approx}, we can obtain a relation between approximate equilibria as well.
\begin{lemma}\label{lem:approx}
Any $\rho$-approximate pure Nash equilibrium for a $\Psi$-game of degree $d$ is a $d!\rho$-approximate pure Nash equilibrium for the corresponding weighted congestion game with polynomial latencies.
\end{lemma}

Since pure Nash equilibria always exist in $\Psi$-games, the last statement (applied with $\rho=1$) implies the following.

\begin{theorem}
Every weighted congestion game with polynomial latency functions of maximum degree $d$ has a $d!$-approximate pure Nash equilibrium.
\end{theorem}

\smallskip\noindent{\bf Subgames and partial potentials.} We now define restrictions of the potential function of $\Psi$-games. Given a state $S$ and a set of players $A\subseteq {\cal N}$, we denote by $N_e^A(S)$ the multiset of the weights of players in $A$ that use resource $e$ in $S$. Then, we define
$$\Phi^A(S) = \sum_e{\sum_{k=0}^d{\frac{a_{e,k}}{k+1}\Psi_{k+1}(N^A_e(S))}}.$$
We can think of $\Phi^A$ as the potential of a subgame in which only the players of $A$ participate.

We also use the notion of the {\em partial} potential to account for the contribution of subsets of players to the potential function. Consider sets of players $A$ and $B$ with $B\subseteq A\subseteq {\cal N}$. Then, the $B$-partial potential of the subgame among the players in $A$ is defined as
$$\Phi_B^A(S) = \Phi^A(S) - \Phi^{A\setminus B}(S).$$
When $A={\cal N}$, we remove the superscript from partial potentials, i.e., $\Phi_B(S) = \Phi_B^{\cal N}(S)$. Also, when $B$ is a singleton containing player $u$, we simplify the notation of the partial potential to $\Phi_u^A(S)$. Furthermore, observe that $\Phi^A_A(S)=\Phi^A(S)$.

The next four claims present basic properties of partial potentials.
\begin{claim}\label{claim:partial-potential-bound}
Let $S$ be a state of a $\Psi$-game and let $B\subseteq A\subseteq {\cal N}$. Then, $\Phi_B^A(S) \leq \Phi_B(S)$.
\end{claim}

\begin{claim}\label{claim:equal-partial-potential}
Let $A\subseteq {\cal N}$ be a set of players and let $S$ and $S'$ be states such that each player in $A$ uses the same strategy in $S$ and $S'$. Then, for every set of players $B\subseteq A$, $\Phi^A_B(S)=\Phi^A_B(S')$.
\end{claim}

\begin{claim}\label{claim:u-partial-potential}
Let $S$ be a state of a $\Psi$-game and let $u$ be a player. Then, $\Phi_u(S) = \hat{c}_u(S)$.
\end{claim}

\begin{claim}\label{claim:partial-potential}
Let $u$ be a player and $A\subseteq {\cal N}$ a set of players that contains $u$. Then, for any two states $S$ and $S'$ that differ only in the strategy of player $u$, it holds that
$\Phi_A(S)-\Phi_A(S') = \hat{c}_u(S)-\hat{c}_u(S')$.
\end{claim}
In particular, Claim \ref{claim:partial-potential} implies that the $A$-partial potential can be thought of as a potential function defined over all states in which each player in ${\cal N}\setminus A$ uses the same strategy.

We proceed with the following interesting property that shows that the potential function of $\Psi$-games is cost-revealing. It also implies that the potential of a state lower-bounds the total cost of all players.
\begin{lemma}\label{lem:cost-vs-potential-d}
For every state $S$ of a $\Psi$-game and any set of players $A\subseteq {\cal N}$, it holds that $\Phi_A(S)\leq \sum_{u\in A}{\hat{c}_u(S)}$.
\end{lemma}

\medskip\noindent{\bf The stretch of the potential function.} An important quantity for our purposes is the {\em stretch} of the potential function of $\Psi$-games; a general definition that applies to every potential game follows.
\begin{dfn}
Consider a potential game with a positive potential function $\Phi$ and let $S^*$ be the state of minimum potential. The $\rho$-stretch of the potential function of the game is the maximum over all $\rho$-approximate pure Nash equilibria $S$ of the ratio $\Phi(S)/\Phi(S^*)$.
\end{dfn}

The next two statements provide bounds on the $\rho$-stretch of the potential function of $\Psi$-games of degree $1$ (i.e., linear weighted congestion games) and $d\geq 2$, respectively.

\begin{lemma}\label{lem:stretch-linear}
For every $\rho\in [1,11/10]$, the $\rho$-stretch of the potential function of a linear weighted congestion game is at most $\frac{3+\sqrt{5}}{2} +6(\rho-1)$.
\end{lemma}

\begin{lemma}\label{lem:stretch-d}
The $\rho$-stretch of the potential function of a $\Psi$-game of degree $d\geq 2$ is at most $\rho(\rho+1)^d(d+1)^{d+1}$.
\end{lemma}

In the rest of the paper, we denote by $\theta_d(\rho)$ the upper bounds on the $\rho$-stretch given by Lemmas \ref{lem:stretch-linear} and \ref{lem:stretch-d}, namely $\theta_1(\rho)=\frac{3+\sqrt{5}}{2}+6(\rho-1)$ and $\theta_d(\rho) = \rho(\rho+1)^d(d+1)^{d+1}$. The next lemma extends these bounds to partial potentials.

\begin{lemma}\label{lem:stretch-partial}
Consider a $\Psi$-game of degree $d$ and a state $S$ which is a $\rho$-approximate pure Nash equilibrium for a set of players $R\subseteq {\cal N}$. Then, $\Phi_R(S) \leq \theta_d(\rho) \Phi_R(S^*)$ for any state $S^*$ such that each player in ${\cal N}\setminus R$ uses the same strategy in $S$ and $S^*$.
\end{lemma}

\section{The algorithm}\label{sec:algo}
In this section we describe our algorithm (Algorithm 1; see the table below). The algorithm takes as input a $\Psi$-game ${\cal G}$ of degree $d$ with $n$ players, an arbitrary initial state $S$ of the game, and a small positive parameter $\gamma$. It produces as output a state of ${\cal G}$. The algorithm starts by initializing its parameters, namely $\hat{c}_{\max}$, $\hat{c}_{\min}$, $m$, $g$, $q$, and $p$ (lines 1-6). It first computes the minimum possible cost $\hat{c}_{\min}$ among all players and the maximum cost $\hat{c}_{\max}$ experienced by players in the initial state $S$. Then, it sets the parameter $m$ equal to $\log{\left(\hat{c}_{\max}/\hat{c}_{\min}\right)}$; in this way, $m$ is polynomial in the number of bits in the representation of the game (i.e., polynomial in the number of bits necessary to store the parameters $a_{e,k}$ and the weights of the players). Then, the parameter $q$ is set close to $1$ (namely, $q=1+\gamma$) and parameter $p$ is set close to $\theta_d(q)$ (namely, $p=\left(\frac{1}{\theta_d(q)} - 2\gamma\right)^{-1}$). Recall that $\theta_d(q)$ is the bound on the $q$-stretch of the potential function of $\Psi$-games of degree $d$ in the statements of Lemmas \ref{lem:stretch-linear} (for $d=1$) and \ref{lem:stretch-d} (for $d\geq 2$).

\IncMargin{3em}
\RestyleAlgo{boxed}
\LinesNumbered
\begin{algorithm}
\SetKwData{Left}{left}\SetKwData{This}{this}\SetKwData{Up}{up}
\SetKwFunction{Union}{Union}\SetKwFunction{FindCompress}{FindCompress}
\SetKwInOut{Input}{input}\SetKwInOut{Output}{output}

\Input{A $\Psi$-game ${\cal G}$ of degree $d$ with a set ${\cal N}$ of $n$ players, an arbitrary initial state $S$, and $\gamma>0$ with $\gamma\in (0,1/10]$ if $d=1$ and $\gamma\in (0,\frac{1}{8\theta_d(2)}]$, otherwise}
\Output{A state of ${\cal G}$}
   $\hat{c}_{\min}\leftarrow\min_{u\in {\cal N}}{\hat{c}_u(\vo_{-u},\BR_u(\vo))}$\;
   $\hat{c}_{\max}\leftarrow \max_{u\in {\cal N}}{\hat{c}_u(S)}$\;
   $m\leftarrow\log{\left(\hat{c}_{\max}/\hat{c}_{\min}\right)}$\;
   $g\leftarrow2\left(1+m(1+\gamma^{-1})\right)^{d}d^{d}n\gamma^{-3}$\;
   $q\leftarrow1+\gamma$\;
   $p \leftarrow \left(\frac{1}{\theta_d(q)} - 2\gamma\right)^{-1}$\; \label{alg:step2}
   \lFor{$i\leftarrow 0$ \KwTo $m$ \label{a}}{$b_i \leftarrow \hat{c}_{\max}g^{-i}$\;} \label{alg:step3}
   \While{there exists a player $u\in {\cal N}$ such that $\hat{c}_u(S)\in [b_1,+\infty)$ and $\hat{c}_u(S_{-u},\BR_u(S))<\hat{c}_u(S)/q$}
   {
   $S\leftarrow (S_{-u},\BR_u(S))$\;
   }
   $F \leftarrow \emptyset$\; \label{alg:step1}
   \For{phase $i\leftarrow 1$ \KwTo $m-1$ \label{main}}{
         \While{there exists a player $u\in {\cal N}\setminus F$ such that either $\hat{c}_u(S) \in [b_{i}, +\infty)$ and $\hat{c}_u(S_{-u},\BR_u(S))<\hat{c}_u(S)/p$ or $\hat{c}_u(S) \in [b_{i+1}, b_{i})$ and $\hat{c}_u(S_{-u},\BR_u(S))<\hat{c}_u(S)/q$}
         {
         $S\leftarrow (S_{-u},\BR_u(S))$\;
         }
         $F \leftarrow F \cup \{u \in {\cal N}\setminus F:\hat{c}_u(S) \in [b_{i}, +\infty)\}$\;
   }
\caption{Computing approximate equilibria in $\Psi$-games.}\label{alg}
\end{algorithm}
\DecMargin{3em}

Then, the algorithm runs a sequence of phases; within each phase, it coordinates best-response moves of the players. This process starts (line 7) by computing a decreasing sequence of boundaries $b_0$, $b_1$, $b_2$, ..., $b_m$ that will be used to define the sets of players that are considered to move within each phase. Then, it executes phase $0$ (lines 8-10). During this phase, as long as there are players of cost at least $b_1$ that have a $q$-move, they play a best-response strategy. Hence, after the end of the phase, all players with cost higher than $b_1$ are in a $q$-approximate equilibrium. Then, the algorithm uses set $F$ to keep the players whose strategies have been irrevocably decided; $F$ is initialized to $\emptyset$ in line 11. Phases $1$ to $m-1$ (lines 12-17) constitute the heart of our algorithm. During each such phase $i$, the algorithm repeatedly checks whether, in the current state, there is a player that either has cost higher than $b_i$ that has a $p$-move or her cost is in $[b_{i+1},b_i)$ and has a $q$-move. While such a player is found, she deviates to her best-response strategy. The phase terminates when no such player exists and the algorithm irrevocably decides the strategy of the players that have cost at least $b_i$. These players are included in set $F$; at this point, they are guaranteed to be at a $p$-approximate equilibrium. Subsequent moves by other players may either increase their cost or decrease the cost they could experience by deviating to another strategy. As we will show, these changes are not significant and each player will still be at an almost $p$-approximate equilibrium at the end of all phases. The fact that plays a crucial role towards proving such a claim is that, at the end of each phase $i$, any player with cost in $[b_{i+1},b_i)$ is guaranteed to be in a $q$-approximate equilibrium. Note that $b_m\leq \hat{c}_{\min}$ and, eventually, all players will be included in set $F$.

We remark that the sequence of the phases is similar to the one in our algorithm for unweighted congestion games with polynomial latency functions of constant degree $d$ in \cite{CFGS11}. However, there is an important difference. In that context, each player is considered to move during only two consecutive phases; these phases are defined statically based only on the characteristics of the particular player. The main reason that allows this is that the cost that a player may experience by following a specific strategy may change by at most a polynomial factor (namely, at most $n^d$) during the execution of the algorithm. This is not the case in the context of $\Psi$-games since the fact that the cost of a player depends on the weights of the other players does not satisfy this polynomial relation. So, in the current algorithm, the players that are considered to move within each phase are decided {\em dynamically} based on the cost they experience during a phase. In this way, a player may (be considered to) move in many different phases.

Below, we will prove the following statement.

\begin{theorem}\label{thm:main}
Algorithm 1 computes a $\hat\rho_d$-approximate equilibrium for every $\Psi$-game of constant degree $d$, where $\hat\rho_1=\frac{3+\sqrt{5}}{2}+O(\gamma)$ and $\hat\rho_d \in d^{d+o(d)}$. The running time is polynomial in $\gamma^{-1}$ and in the number of bits in the representation of the game.
\end{theorem}

Combined with Lemma \ref{lem:approx}, Theorem \ref{thm:main} immediately yields the following result for weighted congestion games.
\begin{theorem}
When Algorithm 1 is applied to the $\Psi$-game corresponding to a weighted congestion game with polynomial latency functions of constant degree $d$, it computes a state which is a $\rho_d$-approximate equilibrium for the latter, where $\rho_1=\frac{3+\sqrt{5}}{2}+O(\gamma)$ and $\rho_d \in d^{2d+o(d)}$ for $d\geq 2$.
\end{theorem}

%\section{Analysis}\label{sec:analysis}
The rest of this section is devoted to proving Theorem \ref{thm:main}. Throughout the section we consider the application of the algorithm on a $\Psi$-game of degree $d$ and denote by $S^i$ the state computed by the algorithm after the execution of phase $i$ for $i=0, 1, ..., m-1$. Also, we use $R_i$ to denote the set of players that make at least one move during phase $i$. Our arguments are split in three parts. First, we present a key property maintained by our algorithm stating that the $R_i$-partial potential is small when the phase $i\geq 1$ starts. Then, we use this fact together with the parameters of the algorithm to prove that the running time is polynomial. The proof of the approximation guarantee follows. Recall that the players whose strategies are irrevocably decided during phase $j\geq 1$ are at a $p$-approximate equilibrium at the end of the phase. The purpose of the third part of the proof is to show that for each such player, neither her cost increases significantly nor the cost she would experience by deviating to another strategy decreases significantly after phase $j$. Hence, the approximation guarantee in the final state computed by the algorithm is slightly higher than $p$.

We remark that the analysis follows the same general steps as in our recent paper on unweighted congestion games \cite{CFGS11}. However, due to the definition of $\Psi$-games and the dependency of players' cost on the weights, different and significantly more involved arguments are required, especially in the first and third step.

The key property maintained by our algorithm is the following.

\begin{lemma}\label{lem:potential-bound-per-phase}
For every phase $i\geq 1$, it holds that $\Phi_{R_i}(S^{i-1}) \leq \gamma^{-1} n b_{i}$.
\end{lemma}

We will now use Lemma \ref{lem:potential-bound-per-phase} and the properties of $\Psi$-games to prove that the algorithm terminates quickly.

\begin{lemma}\label{lem:complexity}
The algorithm terminates after a number of steps that is polynomial in the number of bits in the representation of the game and $\gamma^{-1}$.
\end{lemma}

\begin{proof}
Clearly, if the number of strategies is polynomial in the number of resources, computing a best-response strategy for a player $u$ can be trivially performed in polynomial time (by the definition of $\hat{c}_u$). This is also the case for weighted congestion games in networks (where the number of strategies of a player can be exponential) using a shortest path computation. So, it remains to bound the total number of player moves.

At the initial state, the total cost of the players and, consequently (by Lemma \ref{lem:cost-vs-potential-d}), its potential is at most $n\hat{c}_{\max}$. Each of the players that move during phase $0$ decreases her cost and, consequently (by Theorem \ref{thm:psi-potential}), the potential by at least $(q-1)b_1=\gamma g^{-1} \hat{c}_{\max}$. Hence, the total number of moves in phase $0$ is at most $n \gamma^{-1} g$. For $i\geq 1$, we have $\Phi_{R_i}(S^i)\leq n b_i \gamma^{-1}$ (by Lemma \ref{lem:potential-bound-per-phase}). Each of the players in $R_i$ that move during phase $i$ decreases her cost and, consequently (by Claim \ref{claim:partial-potential}), the $R_i$-partial potential by at least $(q-1)b_{i+1} = b_i g^{-1} \gamma$. Hence, phase $i$ completes after at most $n g \gamma^{-2}$ moves. In total, we have at most $m n g \gamma^{-2}$ moves. The theorem follows by observing that $g$ depends polynomially on $m$, $n$, and $\gamma^{-1}$.
\end{proof}

It remains to prove that our algorithm computes approximate equilibria. Our proofs will exploit Lemma \ref{lem:potential-bound-per-phase} as well as the following lemma which relates the cost of a player in a state to the partial potential of two different subgames.
\begin{lemma}\label{lem:effect}
Consider a $\Psi$-game of degree $d$, a player $u$ and a set of players $R\subseteq {\cal N}\setminus \{u\}$. Then, for every state $S$ and every $\epsilon>0$, it holds that
$$\hat{c}_u(S) \leq (1+\epsilon) \Phi_u^{{\cal N}\setminus R}(S)+\xi_\epsilon \Phi_R^{{\cal N}\setminus \{u\}}(S),$$
where $\xi_\epsilon=(1+1/\epsilon)^{d}d^{d}-1$.
\end{lemma}

Using Lemmas \ref{lem:potential-bound-per-phase} and \ref{lem:effect}, we will show that neither the cost of a player increases significantly after the phase at the end of which her strategy was irrevocably decided (in Lemma \ref{lem:cost-increases-only-slightly}), nor the cost she would experience by deviating to another strategy decreases significantly (in Lemma \ref{lem:deviation-cost-decreases-only-slightly}).

\begin{lemma}\label{lem:cost-increases-only-slightly}
Let $u$ be a player whose strategy was irrevocably decided at phase $j$. Then, $\hat{c}_u(S^{m-1}) \leq (1+2\gamma)\hat{c}_u(S^j)$.
\end{lemma}

\begin{proof}
For every $i>j$ and $\epsilon>0$, we apply Lemma \ref{lem:effect} for strategy $S^i$, player $u$, and the set of players $R_i$ that move during phase $i$ to obtain
\begin{eqnarray*}
\hat{c}_u(S^i) &\leq & (1+\epsilon) \Phi_u^{{\cal N}\setminus R_i}(S^i)+\xi_\epsilon\Phi_{R_i}^{{\cal N}\setminus\{u\}}(S^i)\\
&=& (1+\epsilon) \Phi_u^{{\cal N}\setminus R_i}(S^{i-1})+\xi_\epsilon\Phi_{R_i}^{{\cal N}\setminus\{u\}}(S^i)\\
&\leq & (1+\epsilon) \Phi_u(S^{i-1})+\xi_\epsilon\Phi_{R_i}(S^i)\\
&\leq& (1+\epsilon) \hat{c}_u(S^{i-1})+\xi_\epsilon\Phi_{R_i}(S^{i-1}).
\end{eqnarray*}
The equality holds by Claim \ref{claim:equal-partial-potential} since the players in ${\cal N} \setminus R_i$ do not move during phase $i$. The second inequality follows by Claim \ref{claim:partial-potential-bound}. The last one follows by Claim \ref{claim:u-partial-potential} and since the $R_i$-partial potential decreases during phase $i$.

We now set $\epsilon=(1+\gamma)^{1/m}-1$. This implies that $(1+\epsilon)^m=1+\gamma$. Also, by Claim \ref{claim:concave} (in Appendix \ref{sec:app:two-ineq}), we get $\epsilon \geq \frac{\gamma}{m}(1+\gamma)^{1/m-1}\geq (m(1+\gamma^{-1}))^{-1}$ and, by the definition of the parameters $g$ and $\gamma$, $\xi_\epsilon= (1+m(1+\gamma^{-1}))^d d^d -1 \leq \frac{g\gamma^3}{2n}\leq \frac{g\gamma}{2(1+\gamma^{-1})n}$. Using the above inequality together with these observations, we obtain
\begin{eqnarray*}
\hat{c}_u(S^{m-1}) &\leq & (1+\epsilon)^{m-1-j} \hat{c}_u(S^{j})+\xi_\epsilon\sum_{i=j+1}^{m-1}{(1+\epsilon)^{m-1-i}\Phi_{R_i}(S^{i-1})}\\
&\leq & (1+\epsilon)^{m} \hat{c}_u(S^{j})+(1+\epsilon)^{m}\xi_\epsilon\sum_{i=j+1}^{m-1}{\Phi_{R_i}(S^{i-1})}\\
&\leq& (1+\gamma)\hat{c}_u(S^{j})+(1+\gamma)\xi_\epsilon\sum_{i=j+1}^{m-1}{nb_i \gamma^{-1}}\\
&= & (1+\gamma) \hat{c}_u(S^{j})+(1+\gamma^{-1})\xi_\epsilon n b_j\sum_{i=1}^{m-1-j}{g^{-i}}\\
&\leq & (1+\gamma) \hat{c}_u(S^{j})+2(1+\gamma^{-1})\xi_\epsilon n b_jg^{-1}\\
&\leq & (1+\gamma) \hat{c}_u(S^{j})+\gamma b_j\\
&\leq & (1+2\gamma)\hat{c}_u(S^j).
\end{eqnarray*}
The second inequality is obvious, the third one follows by Lemma \ref{lem:potential-bound-per-phase} and by the relation between $\epsilon$ and $\gamma$, the equality follows by the definition of $b_i$, the fourth inequality follows since $g\geq 2$ which implies that $\sum_{i\geq 1}{g^{-i}}\leq 2g^{-1}$, the fifth one follows by our observation about $\xi_\epsilon$ above, and the last one follows since, by the definition of the algorithm, the fact that the strategy of player $u$ is irrevocably decided at phase $j$ implies that $\hat{c}_u(S^j)\geq b_j$.
\end{proof}

\begin{lemma}\label{lem:deviation-cost-decreases-only-slightly}
Let $u$ be a player whose strategy was irrevocably decided at phase $j$ and let $s'_u$ be any of her strategies. Then, $\hat{c}_u(S^{m-1}_{-u},s'_u) \geq (1-2\gamma)\hat{c}_u(S^j_{-u},s'_u)$.
\end{lemma}

We are now ready to use the last two lemmas in order to prove the approximation guarantee of the algorithm. This will complete the proof of Theorem \ref{thm:main}.
\begin{lemma}\label{lem:apx-bound}
Given a $\Psi$-game of degree $d$, the algorithm computes a $\hat\rho_d$-approximate equilibrium with $\hat\rho_1\leq \frac{3+\sqrt{5}}{2}+O(\gamma)$ and $\hat\rho_d\leq d^{d+o(d)}$.
\end{lemma}

\section{Conclusions and open problems}\label{sec:open}
Due to lack of space, the modification of Algorithm 1 that yields our structural result for weighted congestion games with superlinear latency functions is presented in Appendix \ref{sec:modified-alg}. 

Our work reveals interesting open problems. The obvious one is whether approximate equilibria with a better approximation guarantee can be computed in polynomial time. We believe that our techniques have reached their limits for linear weighted congestion games. However, in the case of superlinear latency functions, approximations of weighted congestion games by potential games different than $\Psi$-games might yield improved (existential or algorithmic) approximation guarantees. On the conceptual level, it is interesting to further explore applications of approximations of non-potential games by potential ones like the one we have exploited in the current paper.

%\small

\newpage\appendix
\section{Two technical inequalities}\label{sec:app:two-ineq}
The following technical inequalities are extensively used in our proofs and are included here for easy reference.

\begin{lemma}[Minkowski inequality]\label{lem:minkowski}
$\sum_{t=1}^s{(\alpha_t+\beta_t)^k}\leq \left(\left(\sum_{t=1}^s{\alpha_t^k}\right)^{1/k}+\left(\sum_{t=1}^s{\beta_t^k}\right)^{1/k}\right)^k$, for any integer $k\geq 1$ and $\alpha_t,\beta_t\geq 0$.
\end{lemma}

\begin{claim}\label{claim:concave}
For every $\alpha\in (0,1)$ and $z>1$, it holds that $z^{\alpha}-1 \geq \alpha (z-1) z^{\alpha-1}$.
\end{claim}

\begin{proof}
The function $h(x)=x^{\alpha}$ is concave in $[1,+\infty)$. This means that, for every $z>1$, the line connecting points $(1,1)$ and $(z, h(z))$ has slope higher than the derivative of $h$ at point $z$, i.e., $\frac{z^{\alpha}-1}{z-1} \geq \alpha z^{\alpha-1}$. Equivalently, $z^{\alpha}-1 \geq \alpha (z-1) z^{\alpha-1}$.
\end{proof}

\section{Omitted proofs from Section \ref{sec:psi-games}}
The following lemma is proved in (the full version of) \cite{C09} and is extensively used in our proofs.

\begin{lemma}\label{lem:properties}
For any integer $k\geq 1$, any finite multi-set of non-negative reals $A$, and any non-negative real $b$ the following hold:
\[\begin{array}{l l}
\mbox{a. } L(A)^k \leq \Psi_k(A) \leq k! L(A)^k & \mbox{d. } \Psi_k(A\cup\{b\}) -\Psi_k(A) = k b\Psi_{k-1}(A\cup \{b\})\\
\mbox{b. } \Psi_{k-1}(A)^{k} \leq \Psi_{k}(A)^{k-1} & \mbox{e. } \Psi_k(A) \leq k\Psi_1(A)\Psi_{k-1}(A)\\
\mbox{c. } \Psi_k(A\cup\{b\}) = \sum_{t=0}^k{\frac{k!}{(k-t)!}b^t\Psi_{k-t}(A)} & \mbox{f. } \Psi_k(A\cup \{b\}) \leq \left(\Psi_k(\{b\})^{1/k}+\Psi_k(A)^{1/k}\right)^k
\end{array}\]
\end{lemma}

\subsection{Proof of Theorem \ref{thm:psi-potential}}
Consider a player $u$, a state $S$ in which $u$ plays strategy $s_u$ and state $(S_{-u},s'_u)$ where $u$ has deviated to strategy $s'_u$. Using the definition of the potential function, we have
\begin{eqnarray*}
\Phi(S)-\Phi(S_{-u},s'_u) &=& \sum_e{\sum_{k=0}^d{\frac{a_{e,k}}{k+1}\Psi_{k+1}(N_e(S))}} - \sum_e{\sum_{k=0}^d{\frac{a_{e,k}}{k+1}\Psi_{k+1}(N_e(S_{-u},s'_u))}}\\
&=& \sum_{e\in s_u \setminus s'_u}{\sum_{k=0}^d{\frac{a_{e,k}}{k+1}\left(\Psi_{k+1}(N_e(S))-\Psi_{k+1}(N_e(S_{-u},s'_u))\right)}}\\
& & +\sum_{e\in s'_u \setminus s_u}{\sum_{k=0}^d{\frac{a_{e,k}}{k+1}\left(\Psi_{k+1}(N_e(S))-\Psi_{k+1}(N_e(S_{-u},s'_u))\right)}}\\
&=& \sum_{e\in s_u \setminus s'_u}{\sum_{k=0}^d{a_{e,k}w_u\Psi_{k}(N_e(S))}} - \sum_{e\in s'_u \setminus s_u}{\sum_{k=0}^d{a_{e,k}w_u\Psi_{k}(N_e(S_{-u},s'_u))}}\\
&=& w_u\sum_{e\in s_u}{\sum_{k=0}^d{a_{e,k}\Psi_{k}(N_e(S))}} - w_u\sum_{e\in s'_u}{\sum_{k=0}^d{a_{e,k}\Psi_{k}(N_e(S_{-u},s'_u))}}\\
&=& \hat{c}_u(S) - \hat{c}_u(S_{-u},s'_u).
\end{eqnarray*}
The third equality follows by Lemma \ref{lem:properties}d and the facts that $N_e(S) = N_e(S_{-u},s'_u)\cup \{w_u\}$ for every resource $e\in s_u\setminus s'_u$ and $N_e(S_{-u},s'_u) = N_e(S)\cup \{w_u\}$ for every resource $e\in s'_u\setminus s_u$. The last equality follows by the definition of $\hat{c}_u$.
\qed

\subsection{Proof of Claim \ref{claim:approx}}
We will use Lemma \ref{lem:properties}a and the definitions of $c_u(S)$ and $\hat{c}_u(S)$. Let $s_u$ be the strategy of player $u$ at state $S$. Using the first inequality of Lemma \ref{lem:properties}a, we have
\begin{eqnarray*}
c_u(S) &=& w_u\sum_{e\in s_u}{\sum_{k=0}^d{a_{e,k}L(N_e(S))^k}} \leq w_u\sum_{e\in s_u}{\sum_{k=0}^d{a_{e,k}\Psi_k(N_e(S))}}= \hat{c}_u(S).
\end{eqnarray*}
Also, using the second inequality in Lemma \ref{lem:properties}a, we have
\begin{eqnarray*}
\hat{c}_u(S) &=& w_u\sum_{e\in s_u}{\sum_{k=0}^d{a_{e,k}\Psi_k(N_e(S))}} \leq w_u\sum_{e\in s_u}{\sum_{k=0}^d{a_{e,k}k!L(N_e(S))^k}}\leq d! w_u\sum_{e\in s_u}{\sum_{k=0}^d{a_{e,k}L(N_e(S))^k}}\\
&=& d! c_u(S).
\end{eqnarray*}
\qed

\subsection{Proof of Lemma \ref{lem:approx}}
Let $S$ be $\rho$-approximate equilibrium for a $\Psi$-game of degree $d$, $u$ a player and $s'_u$ a strategy of $u$ different than her strategy $s_u$ in $S$. Using the $\rho$-approximate equilibrium condition for player $u$ and Claim \ref{claim:approx}, we have
\begin{eqnarray*}
c_u(S) &\leq & \hat{c}_u(S) \leq  \rho \hat{c}_u(S_{-u},s'_u)= d! \rho \cdot c_u(S_{-u},s'_u).
\end{eqnarray*}
\qed

\subsection{Proof of Claim \ref{claim:partial-potential-bound}}
Let $k\geq 1$ be an integer and consider a resource $e$ which is used by at least one player of $B$ in $S$. By the definition of $\Psi_k$, observe that $\Psi_k(N^A_e(S))-\Psi_k(N^{A\setminus B}_e(S))$ is equal to $k!$ times the sum of all  monomials of degree $k$ among the elements of $N^A_e(S)$ that contain at least one element in $N^B_e(S)$. Similarly, $\Psi_k(N_e(S))-\Psi_k(N^{{\cal N}\setminus B}_e(S))$ is equal to $k!$ times the sum of all  monomials of degree $k$ among the elements of $N_e(S)$ that contain at least one element in $N^B_e(S)$. Since $N^A_e(S)\subseteq N_e(S)$, we have that
\begin{eqnarray*}
\Psi_k(N^A_e(S))-\Psi_k(N^{A\setminus B}_e(S)) &\leq & \Psi_k(N_e(S))-\Psi_k(N^{{\cal N}\setminus B}_e(S)).
\end{eqnarray*}
The inequality holds trivially (with equality) if no player from $B$ uses resource $e$ in $S$. Using this inequality and the definition of the partial potential, we have
\begin{eqnarray*}
\Phi^A_B(S) &=& \Phi^A(S)-\Phi^{A\setminus B}(S) = \sum_e{\sum_{k=0}^d{\frac{a_{e,k}}{k+1}\left(\Psi_{k+1}(N^A_e(S))-\Psi_{k+1}(N^{A\setminus B}_e(S))\right)}}\\
&\leq & \sum_e{\sum_{k=0}^d{\frac{a_{e,k}}{k+1}\left(\Psi_{k+1}(N_e(S))-\Psi_{k+1}(N^{{\cal N}\setminus B}_e(S))\right)}} = \Phi(S)-\Phi^{{\cal N}\setminus B}(S)\\
&=& \Phi_B(S).
\end{eqnarray*}
\qed

\subsection{Proof of Claim \ref{claim:equal-partial-potential}}
Observe that $N^{A'}_e(S)=N^{A'}_e(S')$ for each resource $e$ and any $A'\subseteq A$. By the definition of the potential of the subgame among the players of $A'$, we have $\Phi^{A'}(S)=\Phi^{A'}(S')$. Then, by the definition of the partial potential, $\Phi^A_B(S)=\Phi^A(S)-\Phi^{A\setminus B}(S) = \Phi^A(S')-\Phi^{A\setminus B}(S')=\Phi^A_B(S')$.
\qed

\subsection{Proof of Claim \ref{claim:u-partial-potential}}
Let $s_u$ be the strategy of player $u$ in $S$. We use the definition of the partial potential, the definitions of the potential for the original game and the subgame among the players in ${\cal N}\setminus \{u\}$, Lemma \ref{lem:properties}d, and the definition of $\hat{c}_u(S)$ to obtain
\begin{eqnarray*}
\Phi_u(S) &=& \Phi(S)-\Phi^{{\cal N}\setminus \{u\}}(S) = \sum_e{\sum_{k=0}^d{\frac{a_{e,k}}{k+1}\left(\Psi_{k+1}(N_e(S))-\Psi_{k+1}(N^{{\cal N}\setminus \{u\}}_e(S))\right)}}\\
&=& \sum_{e\in s_u}{\sum_{k=0}^d{\frac{a_{e,k}}{k+1}\left(\Psi_{k+1}(N_e(S))-\Psi_{k+1}(N^{{\cal N}\setminus \{u\}}_e(S))\right)}} = w_u \sum_{e\in s_u}{\sum_{k=0}^d{a_{e,k}\Psi_{k}(N_e(S))}}\\
&=& \hat{c}_u(S).
\end{eqnarray*}
\qed

\subsection{Proof of Claim \ref{claim:partial-potential}}
We have
\begin{eqnarray*}
\Phi_A(S)-\Phi_A(S') &=& \Phi(S)-\Phi^{{\cal N}\setminus A}(S)-\Phi(S')+\Phi^{{\cal N}\setminus A}(S') = \Phi(S)-\Phi(S') = \hat{c}_u(S)-\hat{c}_u(S').
\end{eqnarray*}
The first equality follows by the definition of the $A$-partial potential, the second one follows by Claim \ref{claim:equal-partial-potential} since each player in ${\cal N}\setminus A$ uses the same strategy in $S$ and $S'$ and the last one follows by Theorem \ref{thm:psi-potential}.
\qed

\subsection{Proof of Lemma \ref{lem:cost-vs-potential-d}}
Let $A=\{u_1, u_2, ..., u_{|A|}\}$. Let $A_0=\emptyset$ and $A_t=\{u_1, ..., u_t\}$ for $t=1, 2, ..., |A|$. Then, using the definition of the partial potential and Claims \ref{claim:partial-potential-bound} and \ref{claim:u-partial-potential}, we have
\begin{eqnarray*}
\Phi_A(S) &=& \Phi(S)-\Phi^{{\cal N}\setminus A}(S) = \sum_{t=1}^{|A|}{\left(\Phi^{{\cal N}\setminus A_{t-1}}(S) - \Phi^{{\cal N}\setminus A_{t}}(S)\right)}\\
&=& \sum_{t=1}^{|A|}{\Phi^{{\cal N}\setminus A_{t-1}}_{u_t}(S)} \leq  \sum_{t=1}^{|A|}{\Phi_{u_t}(S)} = \sum_{u \in A}{\hat{c}_u(S)}.
\end{eqnarray*}
\qed

\subsection{Proof of Lemma \ref{lem:stretch-linear}}
Let $S^*$ be the state of minimum potential and $S$ be a $\rho$-approximate equilibrium. For each player $u$, we denote by $s_u$ and $s^*_u$ the strategies she plays at states $S$ and $S^*$, respectively. Using the $\rho$-approximate equilibrium condition $c_u(S)\leq \rho\cdot c_u(S_{-u},s^*_u)$, the definition of the cost of player $u$, and the definition of function $\Psi_1$, we obtain
\begin{eqnarray*}
\sum_u{c_u(S)} & \leq & \rho w_u\sum_{e\in s^*_u}{\left(a_{e,1}\Psi_1(N_e(S_{-u},s^*_u))+a_{e,0}\right)}\\
& \leq & \rho w_u\sum_{e\in s^*_u}{\left(a_{e,1}\Psi_1(N_e(S)\cup \{w_u\})+a_{e,0}\right)}\\
&=& \rho w_u\sum_{e\in s^*_u}{\left(a_{e,1}\Psi_1(N_e(S))+a_{e,1}w_u+a_{e,0}\right)}.
\end{eqnarray*}
By summing over all players, by exchanging sums, and using the definition of $N_e(S^*)$, we obtain
\begin{eqnarray*}
\sum_u{c_u(S)} &\leq & \rho \sum_u{w_u\sum_{e\in s^*_u}{\left(a_{e,1}\Psi_1(N_e(S))+a_{e,1}w_u+a_{e,0}\right)}}\\
&=& \rho\sum_e{\left(a_{e,1}\Psi_1(N_e(S))\sum_{u:e\in s^*_u}{w_u}+a_{e,1}\sum_{u:e\in s^*_u}{w^2_u}+a_{e,0}\sum_{u:e\in s^*_u}{w_u}\right)}\\
&=& \rho\sum_e{\left(a_{e,1}\Psi_1(N_e(S))\Psi_1(N_e(S^*))+a_{e,1}\sum_{u:e\in s^*_u}{w^2_u}+a_{e,0}\Psi_1(N_e(S^*))\right)}.
\end{eqnarray*}
We now apply the inequality $xy\leq \frac{\sqrt{5}-1}{2(3-\sqrt{5})}y^2+\frac{\sqrt{5}-2}{3-\sqrt{5}}x^2$ that holds for any pair of non-negative $x$ and $y$ on the rightmost part of the above derivation to obtain
\begin{eqnarray*}
& & \sum_u{c_u(S)} \\
&\leq & \rho\sum_e{\left(\frac{\sqrt{5}-1}{2(3-\sqrt{5})}a_{e,1}\Psi_1(N_e(S^*))^2+\frac{\sqrt{5}-2}{3-\sqrt{5}}a_{e,1}\Psi_1(N_e(S))^2+a_{e,1}\sum_{u:e\in s^*_u}{w^2_u}+a_{e,0}\Psi_1(N_e(S^*))\right)}\\
&=& \rho\sum_e{\left(\frac{5-\sqrt{5}}{4(3-\sqrt{5})}a_{e,1}\left(\Psi_1(N_e(S^*))^2+\sum_{u:e\in s^*_u}{w^2_u}\right)+a_{e,0}\Psi_1(N_e(S^*))\right)}\\
& & -\rho\sum_e{\frac{7-3\sqrt{5}}{4(3-\sqrt{5})}a_{e,1}\left(\Psi_1(N_e(S^*))^2-\sum_{u:e\in s^*_u}{w^2_u}\right)}+\frac{\sqrt{5}-2}{3-\sqrt{5}}q\sum_e{a_{e,1}\Psi_1(N_e(S))^2}.
\end{eqnarray*}
Now, observe that $\Psi_1(N_e(S^*))^2 \geq \sum_{u:e\in s^*_u}{w^2_u}$ for every resource $e$. Furthermore, $\Psi_1(N_e(S^*))^2+\sum_{u:e\in s^*_u}{w^2_u}=\Psi_2(N_e(S^*))$. Hence, we have
\begin{eqnarray}\nonumber
\sum_u{c_u(S)} &\leq & \rho\sum_e{\left(\frac{5-\sqrt{5}}{4(3-\sqrt{5})}a_{e,1}\Psi_2(N_e(S^*))+a_{e,0}\Psi_1(N_e(S^*))\right)}+\frac{\sqrt{5}-2}{3-\sqrt{5}}\rho\sum_e{a_{e,1}\Psi_1(N_e(S))^2}\\\nonumber
&\leq & \frac{5-\sqrt{5}}{2(3-\sqrt{5})} \rho\sum_e{\left(\frac{a_{e,1}}{2}\Psi_2(N_e(S^*))+a_{e,0}\Psi_1(N_e(S^*))\right)}+\frac{\sqrt{5}-2}{3-\sqrt{5}}\rho\sum_e{a_{e,1}\Psi_1(N_e(S))^2}\\\label{eq:total-cost-linear}
&= & \frac{5-\sqrt{5}}{2(3-\sqrt{5})}\rho \Phi(S^*)+\frac{\sqrt{5}-2}{3-\sqrt{5}}\rho\sum_e{a_{e,1}\Psi_1(N_e(S))^2}.
\end{eqnarray}

We now use the definition of $\Phi(S)$, the fact that for every player $u$ and resource $e\in s_u$, it holds that $w_u\leq \Psi_1(N_e(S))$, and the definition of the cost of player $u$. We have
\begin{eqnarray*}
\Phi(S) &=& \sum_e{\left(\frac{a_{e,1}}{2}\Psi_2(N_e(S))+a_{e,0}\Psi_1(N_e(S))\right)}\\
&=& \sum_e{\left(\frac{a_{e,1}}{2}\sum_{u:e \in s_u}{\left(w_u \Psi_1(N_e(S))+w^2_u\right)}+a_{e,0}\sum_{u:e\in s_u}{w_u}\right)}\\
&\leq & \sum_e{\left(\frac{a_{e,1}}{2}\sum_{u:e \in s_u}{\left((6-2\sqrt{5})w_u \Psi_1(N_e(S))+(2\sqrt{5}-4)w^2_u\right)}+a_{e,0}\sum_{u:e\in s_u}{w_u}\right)}\\
&=& (3-\sqrt{5})\sum_u{w_u\sum_{e\in s_u}{\left(a_{e,1}\Psi_1(N_e(S))+a_{e,0}\right)}}+(\sqrt{5}-2)\sum_e{a_{e,1}\sum_{u:e\in s_u}{w^2_u}}\\
&& +(\sqrt{5}-2)\sum_e{a_{e,0}\sum_{u:e\in s_u}{w_u}}\\
&=& (3-\sqrt{5})\sum_u{c_u(S)} +(\sqrt{5}-2)\sum_e{a_{e,1}\sum_{u:e\in s_u}{w^2_u}}+(\sqrt{5}-2)\sum_e{a_{e,0}\sum_{u:e\in s_u}{w_u}}.
\end{eqnarray*}
By applying inequality (\ref{eq:total-cost-linear}) to the rightmost part of this derivation, we obtain
\begin{eqnarray*}
\Phi(S) &\leq & \frac{5-\sqrt{5}}{2}\rho \Phi(S^*)+(\sqrt{5}-2)\rho\sum_e{a_{e,1}\Psi_1(N_e(S))^2}+(\sqrt{5}-2)\sum_e{a_{e,1}\sum_{u:e\in s_u}{w^2_u}}\\
&& +(\sqrt{5}-2)\sum_e{a_{e,0}\Psi_1(N_e(S))}\\
&\leq & \frac{5-\sqrt{5}}{2}\rho\Phi(S^*)+(2\sqrt{5}-4)\rho\sum_e{\left(\frac{a_{e,1}}{2}\left(\Psi_1(N_e(S))^2+\sum_{u:e\in s_u}{w^2_u}\right)+a_{e,0}\Psi_1(N_e(S))\right)}\\
&=& \frac{5-\sqrt{5}}{2}\rho\Phi(S^*)+(2\sqrt{5}-4)\rho\sum_e{\left(\frac{a_{e,1}}{2}\Psi_2(N_e(S))+a_{e,0}\Psi_1(N_e(S))\right)}\\
&=& \frac{5-\sqrt{5}}{2}\rho\Phi(S^*)+(2\sqrt{5}-4)\rho\Phi(S).
\end{eqnarray*}
The last inequality implies that $\Phi(S)$ is not larger than $\frac{(5-\sqrt{5})\rho}{2(1-(2\sqrt{5}-4)\rho)} \Phi(S^*)$ which can be easily proved to be at most $\left(\frac{3+\sqrt{5}}{2}+6(\rho-1)\right) \Phi(S^*)$ when $\rho\in [1,11/10]$.
\qed

\subsection{Proof of Lemma \ref{lem:stretch-d}}
Consider a $\rho$-approximate equilibrium $S$ of a $\Psi$-game and let $S^*$ be the state of minimum potential. We denote by $s_u$ and $s^*_u$ the strategy of player $u$ at states $S$ and $S^*$, respectively.

By Lemma \ref{lem:cost-vs-potential-d}, the $\rho$-approximate equilibrium condition $\hat{c}_u(S)\leq \rho \cdot \hat{c}_u(S_{-u},s^*_u)$, and the definition of the potential function, we have
\begin{eqnarray*}
\frac{1}{\rho}\Phi(S) &\leq & \frac{1}{\rho}\sum_u{\hat{c}_u(S)}\\
&\leq & \sum_u{\hat{c}_u(S_{-u},s^*_u)}\\
&=& \sum_u{w_u\sum_{e\in s^*_u}{\sum_{k=0}^d{a_{e,k}\Psi_k(N_e(S_{-u},s^*_u))}}}\\
&=& \sum_e{\sum_{k=0}^d{a_{e,k}\sum_{u:e\in s^*_u}{w_u\Psi_k(N_e(S_{-u},s^*_u))}}}.
\end{eqnarray*}
We now use the fact that $N_e(S_{-u},s^*_u)\subseteq N_e(S)\cup \{w_u\}$, Lemma \ref{lem:properties}c, and the fact that $\Psi_{t+1}(N_e(S^*))\geq (t+1)! \sum_{u:e\in s^*_u}{w^{t+1}_u}$ to obtain
\begin{eqnarray*}
\frac{1}{\rho}\Phi(S) &\leq & \sum_e{\sum_{k=0}^d{a_{e,k}\sum_{u:e\in s^*_u}{w_u\Psi_k(N_e(S) \cup \{w_u\})}}}\\
&=& \sum_e{\sum_{k=0}^d{a_{e,k}\sum_{u:e\in s^*_u}{w_u\sum_{t=0}^k{\frac{k!}{(k-t)!}\Psi_{k-t}(N_e(S)) w^t_u}}}}\\
&=& \sum_e{\sum_{k=0}^d{a_{e,k}\sum_{t=0}^k{\frac{k!}{(k-t)!}\Psi_{k-t}(N_e(S))\sum_{u:e\in s^*_u}{w^{t+1}_u}}}}\\
&\leq & \sum_e{\sum_{k=0}^d{a_{e,k}\sum_{t=0}^k{\frac{k!}{(k-t)!(t+1)!}\Psi_{k-t}(N_e(S))\Psi_{t+1}(N_e(S^*))}}}\\
&=& \sum_e{\sum_{k=0}^{d}{\frac{a_{e,k}}{k+1}\sum_{t=1}^{k+1}{\left(\begin{array}{c}k+1\\t\end{array}\right)\Psi_{k+1-t}(N_e(S))\Psi_{t}(N_e(S^*))}}}.
\end{eqnarray*}
Using Lemma \ref{lem:properties}b (observe that it implies that $\Psi_t(A) \leq \Psi_{k+1}(A)^{\frac{t}{k+1}}$ for any non-negative integer $t\leq k+1$ and multi-set of reals $A$), the binomial theorem, inequality $\alpha^\lambda+\beta^{\lambda} \leq (\alpha+\beta)^{\lambda}$ for every $\alpha,\beta\geq 0$ and $\lambda \geq 1$, and the definition of the potential function, we obtain
\begin{eqnarray*}
\frac{1}{\rho}\Phi(S) &\leq & \sum_e{\sum_{k=0}^{d}{\frac{a_{e,k}}{k+1}\sum_{t=1}^{k+1}{\left(\begin{array}{c}k+1\\t\end{array}\right)\Psi_{k+1}(N_e(S))^\frac{k+1-t}{k+1}\Psi_{k+1}(N_e(S^*))^\frac{t}{k+1}}}}\\
&=& \sum_e{\sum_{k=0}^{d}{\frac{a_{e,k}}{k+1}\left(\left(\Psi_{k+1}(N_e(S))^\frac{1}{k+1}+\Psi_{k+1}(N_e(S^*))^\frac{1}{k+1}\right)^{k+1}-\Psi_{k+1}(N_e(S))\right)}}\\
&\leq & \sum_e{\sum_{k=0}^{d}{\frac{a_{e,k}}{k+1}\left(\Psi_{k+1}(N_e(S))^\frac{1}{d+1}+\Psi_{k+1}(N_e(S^*))^\frac{1}{d+1}\right)^{d+1}}}-\Phi(S).
\end{eqnarray*}
We now apply Minkowski inequality twice on the double sum at the rightmost part of this last inequality and use the definition of the potential function to obtain
\begin{eqnarray*}
(1+1/\rho)\Phi(S) &\leq & \sum_e{\left(\left(\sum_{k=0}^{d}{\frac{a_{e,k}}{k+1}\Psi_{k+1}(N_e(S))}\right)^\frac{1}{d+1}+\left(\sum_{k=0}^d{\frac{a_{e,k}}{k+1}\Psi_{k+1}(N_e(S^*))}\right)^\frac{1}{d+1}\right)^{d+1}}\\
&\leq & \left(\left(\sum_e{\sum_{k=0}^{d}{\frac{a_{e,k}}{k+1}\Psi_{k+1}(N_e(S))}}\right)^\frac{1}{d+1}+\left(\sum_e{\sum_{k=0}^d{\frac{a_{e,k}}{k+1}\Psi_{k+1}(N_e(S^*))}}\right)^\frac{1}{d+1}\right)^{d+1}\\
&=& \left(\left(\Phi(S)\right)^{\frac{1}{d+1}}+\left(\Phi(S^*)\right)^{\frac{1}{d+1}}\right)^{d+1}.
\end{eqnarray*}
The above inequality yields
\begin{eqnarray}\label{eq:potential-d}
\left(\Phi(S)\right)^{\frac{1}{d+1}} &\leq & \frac{1}{(1+1/\rho)^{\frac{1}{d+1}}-1} \left(\Phi(S^*)\right)^{\frac{1}{d+1}}.
\end{eqnarray}
By Claim \ref{claim:concave}, we have $(1+1/\rho)^{\frac{1}{d+1}}-1 \geq \left(\rho^{\frac{1}{d+1}}(\rho+1)^{\frac{d}{d+1}}(d+1)\right)^{-1}$. Using this observation, inequality (\ref{eq:potential-d}) implies that
\begin{eqnarray*}
\Phi(S)&\leq & \rho(\rho+1)^d(d+1)^{d+1}\Phi(S^*)
\end{eqnarray*}
as desired.
\qed

\subsection{Proof of Lemma \ref{lem:stretch-partial}}
In our proof, we will use the property
\begin{eqnarray}\label{eq:psi-property}
\Psi_k(A\cup B) &=& \sum_{t=0}^k{\left(\begin{array}{c}k\\t\end{array}\right)\Psi_{k-t}(A)\Psi_t(B)}
\end{eqnarray}
for every two multi-sets of positive reals $A$ and $B$. To see why (\ref{eq:psi-property}) holds, observe that the product $\Psi_{k-t}(A)\Psi_t(B)$ equals $(k-t)! t!$ times the sum of all products of monomials of degree $k-t$ with elements of $A$ with monomials of degree $t$ with elements of $B$.

Given state $S$ in the original game, we define the $\Psi$-game $\left(R,(w_u)_{u\in R}, (\Sigma_u)_{u\in R}, (\bar{a}_{e,t})_{e\in E, t=0, ..., d}\right)$ with
$$\bar{a}_{e,t} = \sum_{k=t}^d{a_{e,k}\left(\begin{array}{c}k\\t\end{array}\right)\Psi_{k-t}(N^{{\cal N}\setminus R}_e(S))}.$$
Observe that the parameters $\bar{a}_{e,k}$ depend only on the strategies of players in ${\cal N}\setminus R$ in $S$.

Now, given any state $S'$ in the original game, we denote by $\bar{S}'$ the state in the new game in which each player in $R$ uses the strategy she uses in $S'$. We also use the notation $\bar{c}_u$ for the cost of a player $u\in R$ in the new game and $\bar{\Phi}$ for its potential function.

We will first show that $\bar{c}_u(\bar{S}')=\hat{c}_u(S')$ for every state $\bar{S}'$ of the new game such that each player $u\in {\cal N}\setminus R$ uses the same strategy in $S'$ and $S$. Consequently, since state $S$ is a $\rho$-approximate equilibrium for the players in $R$ in the original game, state $\bar{S}$ is a $\rho$-approximate equilibrium in the new game. We have
\begin{eqnarray*}
\bar{c}_u(\bar{S}') &=& w_u\sum_{e\in s_u}{\sum_{t=0}^d{\bar{a}_{e,t}\Psi_t(N_e(\bar{S}'))}} = w_u\sum_{e\in s_u}{\sum_{t=0}^d{\Psi_t(N^R_e(S'))}\sum_{k=t}^d{a_{e,k}\left(\begin{array}{c}k\\t\end{array}\right)\Psi_{k-t}(N^{{\cal N}\setminus R}_e(S))}}\\
&=& w_u\sum_{e\in s_u}{\sum_{k=0}^d{a_{e,k}\sum_{t=0}^k{\left(\begin{array}{c}k\\t\end{array}\right)\Psi_{k-t}(N^{{\cal N}\setminus R}_e(S'))\Psi_t(N^R_e(S'))}}} = w_u\sum_{e\in s_u}{\sum_{k=0}^d{a_{e,k}\Psi_k(N_e(S'))}}\\
&=& \hat{c}_u(S').
\end{eqnarray*}
The first equality follows by the definition of $\bar{c}_u(\bar{S}')$, the second one follows since $N_e(\bar{S}')=N^R_e(S')$ and by the definition of $\bar{a}_{e,k}$, the third one follows by exchanging the sums and since each player in ${\cal N}\setminus R$ use the same strategy in states $S$ and $S'$ (hence, $N_e^{{\cal N}\setminus R}(S)=N_e^{{\cal N}\setminus R}(S')$), the fourth one follows by equality (\ref{eq:psi-property}), and the last one follows by the definition of $\hat{c}_u(S')$.

We now show that $\bar{\Phi}(\bar{S}')=\Phi_R(S')$. We have
\begin{eqnarray*}
\bar{\Phi}(\bar{S}') &=& \sum_e{\sum_{t=0}^d{\frac{\bar{a}_{e,t}}{t+1}\Psi_{t+1}(N_e(\bar{S}'))}}\\
&=& \sum_e{\sum_{t=0}^d{\Psi_{t+1}(N_e^R(S'))\sum_{k=t}^d{a_{e,k}\frac{k!}{(t+1)!(t-k)!}\Psi_{k-t}(N^{{\cal N}\setminus R}_e(S))}}}\\
&=& \sum_e{\sum_{k=0}^d{\frac{a_{e,k}}{k+1}\sum_{t=0}^k{\left(\begin{array}{c}k+1\\t+1\end{array}\right)\Psi_{k-t}(N^{{\cal N}\setminus R}_e(S'))\Psi_{t+1}(N_e^R(S'))}}}\\
&=& \sum_e{\sum_{k=0}^d{\frac{a_{e,k}}{k+1}\sum_{t=1}^{k+1}{\left(\begin{array}{c}k+1\\t\end{array}\right)\Psi_{k+1-t}(N^{{\cal N}\setminus R}_e(S'))\Psi_{t}(N_e^R(S'))}}}\\
&=& \sum_e{\sum_{k=0}^d{\frac{a_{e,k}}{k+1}\left(\sum_{t=0}^{k+1}{\left(\begin{array}{c}k+1\\t\end{array}\right)\Psi_{k+1-t}(N^{{\cal N}\setminus R}_e(S'))\Psi_{t}(N_e^R(S'))}-\Psi_{k+1}(N_e^{{\cal N}\setminus R}(S'))\right)}}\\
&=& \sum_e{\sum_{k=0}^d{\frac{a_{e,k}}{k+1}\Psi_{k+1}(N_e^R(S'))}}-\sum_e{\sum_{k=0}^d{\frac{a_{e,k}}{k+1}\Psi_{k+1}(N_e^{{\cal N}\setminus R}(S'))}}\\
&=& \Phi(S')-\Phi^{{\cal N}\setminus R}(S')\\
&=& \Phi_R(S').
\end{eqnarray*}
The first equality follows by the definition of $\bar{\Phi}(\bar{S}')$, the second one follows since $N_e(\bar{S}')=N^R_e(S')$ and by the definition of $\bar{a}_{e,k}$, the third one follows by exchanging the sums and since each player in ${\cal N}\setminus R$ use the same strategy in states $S$ and $S'$, the fourth one follows by simply changing the counter in the rightmost sum, the fifth one is obvious, the sixth one follows by property (\ref{eq:psi-property}), and the last two ones follow by the definition of the (partial) potentials.

Since the state $\bar{S}$ is a $\rho$-approximate equilibrium for the new game, the bounds on the $\rho$-stretch established in Lemmas \ref{lem:stretch-linear} and \ref{lem:stretch-d} imply that $\bar{\Phi}(\bar{S}) \leq \theta_d(\rho) \bar{\Phi}(\bar{S}^*)$. By our last equality above, we obtain that $\Phi_R(S) \leq \theta_d(\rho) \Phi_R(S^*)$ and the proof is complete.
\qed

\section{Omitted proofs from Section \ref{sec:algo}}

\subsection{Proof of Lemma \ref{lem:potential-bound-per-phase}}
In order to prove the key property maintained by our algorithm, we will need the following lemma which relates the $R_i$-partial potential to the cost they experience when they make their last move within phase $i$.

\begin{lemma}\label{lem:last-moves}
Let $\hat{c}(u)$ denote the cost of player $u\in R_i$ just after making her last move within phase $i\geq 1$. Then,
$$\Phi_{R_i}(S^i) \leq \sum_{u\in R_i}{\hat{c}(u)}.$$
\end{lemma}

\begin{proof}
Rename the players in $R_i$ as $u_1, u_2, ..., u_{|R_i|}$ so that $u_j$ is the $j$-th player that performed her last move within phase $i\geq 1$. Also, denote by $S^{i,j}$ the state in which player $u_j$ performed her last move. Let $R^{|R_i|}_i = \emptyset$ and $R^j_i = \{u_{j+1}, u_{j+2}..., u_{|R_i|}\}$ for $j=0, 1, 2, ..., |R_i|-1$. Then,
\begin{eqnarray*}
\Phi_{R_i}(S^i) &=& \Phi(S^i)-\Phi^{{\cal N}\setminus R_i}(S^i)  =  \sum_{j=1}^{|R_i|}{\left(\Phi^{{\cal N}\setminus R_i^{j}}(S^i)-\Phi^{{\cal N}\setminus R_i^{j-1}}(S^i)\right)} = \sum_{j=1}^{|R_i|}{\Phi_{u_{j}}^{{\cal N}\setminus R_i^j}(S^i)}\\
&= & \sum_{j=1}^{|R_i|}{\Phi_{u_j}^{{\cal N}\setminus R_i^j}(S^{i,j})} \leq  \sum_{j=1}^{|R_i|}{\Phi_{u_j}(S^{i,j})} = \sum_{u\in R_i}{\hat{c}(u)}.
\end{eqnarray*}
The first three inequalities follow by the definition of the partial potential functions and the definition of sets $R_i^j$. The fourth inequality follows by Claim \ref{claim:equal-partial-potential} since players in ${\cal N}\setminus R^j_i$ do not move after state $S^{i,j}$ and until the end of the phase. The inequality follows by Claim \ref{claim:partial-potential-bound} and the last equality follows by Claim \ref{claim:u-partial-potential} and the definition of $\hat{c}(u)$.
\end{proof}

We now proceed to the proof of Lemma \ref{lem:potential-bound-per-phase}. For the sake of contradiction, we assume that $\Phi_{R_i}(S^{i-1}) > \gamma^{-1}n b_i$ and we denote by $P_i$ and $Q_i$ the set of players in $R_i$ whose last move was a $p$-move and $q$-move, respectively. Since each player in $P_i$ decreases her cost by at least $(p-1)\hat{c}(u)$ during her last move within phase $i$ (see Claim \ref{claim:partial-potential}), we have
\begin{eqnarray*}
\Phi_{R_{i}}(S^{i}) &\leq & \Phi_{R_{i}}(S^{i-1})- (p-1)\sum_{u \in P_i} \hat{c}(u).
\end{eqnarray*}
By Lemma \ref{lem:last-moves} and the fact that each player in $Q_i$ experiences a cost of at most $b_i$ when she makes her last move within phase $i$, we have
\begin{eqnarray*}
\sum_{u\in P_i}{\hat{c}(u)} &\geq & \Phi_{R_i}(S^i)-\sum_{u\in Q_i}{\hat{c}(u)} \geq \Phi_{R_i}(S^i)-nb_i.
\end{eqnarray*}
Using the last two inequalities and our assumption, we obtain that
\begin{eqnarray*}
\Phi_{R_i}(S^i) &\leq & \Phi_{R_i}(S^{i-1}) - (p-1) \Phi_{R_i}(S^i)+(p-1) nb_i\\
&< & (1+(p-1)\gamma)\Phi_{R_i}(S^{i-1}) - (p-1) \Phi_{R_i}(S^i)
\end{eqnarray*}
which implies that
\begin{eqnarray*}
\Phi_{R_i}(S^i) &< & \left(\frac{1}{p}+\gamma\right)\Phi_{R_i}(S^{i-1}).
\end{eqnarray*}

Now, consider state $S^{i-1}$ and let $X_i$ and $Y_i$ be the sets of players in $R_i$ with cost at least $b_i$ and smaller than $b_i$, respectively. Notice that, by the definition of phase $i-1$, $S^{i-1}$ is a $q$-approximate equilibrium for the players in $X_i$. We construct a new $\Psi$-game of degree $d$ among the players in $\cal N$ as follows. The new game has all resources of the original game; the parameters $a_{e,k}$ for these resources are the same as in the original game. In addition, the new game has a new resource $e_u$ for each player $u\in Y_i$; the parameters for this resource are $a_{e_u,0}=b_i/w_u$ and $a_{e_u,k}=0$ for $k=1, ..., d$. Each player in ${\cal N} \setminus Y_i$ has the same set of strategies in the two games. The strategy set of player $u\in Y_i$ consists of the strategy $s_u$ she uses in $S^{i-1}$ as well as strategy $s'_u\cup \{e_u\}$ for each strategy $s'_u\not=s_u$ she has in the original game.

Let $\bar{S}^{i-1}$ be the state of the new game in which all players play their strategies in $S^{i-1}$. Clearly, state $\bar{S}^{i-1}$ is a $q$-approximate equilibrium for the players in $X_i$. Also, at state $\bar{S}^{i-1}$, each player $u\in Y_i$ experiences a cost equal to the cost she experiences at state $S^{i-1}$ of the original game, i.e., smaller than $b_i$. In the new game, any deviation of $u$ would include resource $e_u$ and would increase the cost of player $u$ to at least $w_u a_{e_u,0}=b_i$. Hence, $\bar{S}^{i-1}$ is a $q$-approximate equilibrium for the players of $Y_i$ as well. We use $\bar\Phi$ to denote the potential of the new game. Since the players use the same strategies in states $S^{i-1}$ and $\bar{S}^{i-1}$ and the parameters $a_{e,k}$ of the original resources are the same in both games, we have $\bar\Phi_{R_i}(\bar{S}^{i-1}) = \Phi_{R_i}(S^{i-1})$.

Now, let $\bar{S}^i$ be the state in which each player in ${\cal N}\setminus Y_i$ uses her strategy in $S^i$ and the strategies for the players in $Y_i$ are defined as follows. Let $u$ be a player of $Y_i$ and $s'_u$ be the strategy she uses at state $S^i$ of the original game. Her strategy in state $\bar{S}^i$ of the new game is $s'_u\cup \{e_u\}$ if $s'_u\not=s_u$ and $s_u$ otherwise. Observe that, by the definition of the partial potential, we have that the partial potential $\bar\Phi_{R_i}(\bar{S}^i)$ of the new game at state $\bar{S}^i$ is by at most $\sum_{u\in Y_i}{a_{e_u,0} \Psi_1(N_{e_u}(\bar{S}^i))} \leq nb_i$ higher than the partial potential of the original game at state $S^i$ (due to the contribution of the additional resources to the potential value). Hence,
\begin{eqnarray*}
\bar{\Phi}_{R_i}(\bar{S}^i) &\leq & \Phi_{R_i}(S^i)+nb_i <\left(\frac{1}{p}+2\gamma\right) \Phi_{R_i}(S^{i-1}) = \left(\frac{1}{p}+2\gamma\right) \bar{\Phi}_{R_i}(\bar{S}^{i-1}) =  \frac{1}{\theta_d(q)} \bar{\Phi}_{R_i}(\bar{S}^{i-1}).
\end{eqnarray*}
So, we have identified a state $\bar{S}^{i-1}$ of the new game which is a $q$-approximate equilibrium for the players in $R_i$ and another state $\bar{S}^i$ such that the players in ${\cal N}\setminus R_i$ use the same strategies in $\bar{S}^{i-1}$ and $\bar{S}^i$ and $\bar\Phi_{R_i}(\bar{S}^{i-1}) > \theta_d(q) \bar\Phi_{R_i}(\bar{S}^i)$. This contradicts Lemma \ref{lem:stretch-partial} and, subsequently, it also contradicts our assumption $\Phi_{R_i}(S^{i-1})>\gamma^{-1}nb_i$. The lemma follows.
\qed

\subsection{Proof of Lemma \ref{lem:effect}}
In order to prove the lemma, we will need the following technical claim.
\begin{claim}\label{claim:technical}
For any $\alpha,\beta\geq 0$ and integer $d\geq 1$, it holds that $(\alpha+\beta)^{d+1}\leq (1+\epsilon)\alpha^{d+1}+(1+1/\epsilon)^d d^d \beta^{d+1}$.
\end{claim}

\begin{proof}
Consider the function $h(\alpha) = (\alpha+\beta)^{d+1} - (1+\epsilon)\alpha^{d+1}$. By setting its derivative equal to $0$, we obtain that it is maximized for $\alpha=\beta\left((1+\epsilon)^{1/d}-1\right)^{-1}$ to the value $\frac{1+\epsilon}{((1+\epsilon)^{1/d}-1)^d}\beta^{d+1}$. By Claim \ref{claim:concave}, we have that $(1+\epsilon)^{1/d}-1 \geq \frac{\epsilon}{d(1+\epsilon)^{1-1/d}}$. Hence, $h(\alpha) \leq (1+1/\epsilon)^d d^{d} \beta^{d+1}$ as desired.
\end{proof}

Now, let $k$ be an integer such that $1\leq k\leq d+1$, $A$ a multiset of reals, and $b\geq 0$. Using Lemma \ref{lem:properties}f, inequality $\alpha^\lambda+\beta^{\lambda} \leq (\alpha+\beta)^{\lambda}$ for every $\alpha,\beta\geq 0$ and $\lambda \geq 1$, and Claim \ref{claim:technical}, we have
\begin{eqnarray}\nonumber
\Psi_k(A\cup \{b\})-\Psi_k(A) &\leq & \left(\Psi_k(\{b\})^{1/k}+\Psi_k(A)^{1/k}\right)^k - \Psi_k(A)\\\nonumber
&\leq & \left(\Psi_k(\{b\})^{\frac{1}{d+1}}+\Psi_k(A)^{\frac{1}{d+1}}\right)^{d+1} - \Psi_k(A)\\\label{eq:use-technical}
&\leq & (1+\epsilon)\Psi_k(\{b\})+\xi_\epsilon \Psi_k(A).
\end{eqnarray}

Also, let $Q={\cal N}\setminus (R\cup\{u\})$ and define
$$\delta_{e,t} = \sum_{k=\max\{t-1,0\}}^d{\frac{a_{e,k}}{k+1}\left(\begin{array}{c}k+1\\t\end{array}\right)\Psi_{k+1-t}(N_e^Q(S))}$$
for each resource $e$ and $t=0, 1, ..., d+1$. Also, let $P$ be a possibly empty set such that $P\subseteq R\cup\{u\}$. By the definition of function $\Psi_{k+1}$ and by exchanging the sums, we have
\begin{eqnarray}\nonumber
\Phi^{P\cup Q}(S) &=& \sum_e{\sum_{k=0}^d{\frac{a_{e,k}}{k+1}\Psi_{k+1}(N^{P\cup Q}_e(S))}}\\\nonumber
&=& \sum_e{\sum_{k=0}^d{\frac{a_{e,k}}{k+1}\sum_{t=0}^{k+1}\left(\begin{array}{c}k+1\\t\end{array}\right)\Psi_{t}(N^{P}_e(S))\Psi_{k+1-t}(N^Q_e(S))}}\\\nonumber
&=& \sum_e{\sum_{t=0}^{d+1}{\Psi_{t}(N^{P}_e(S))\sum_{k=\max\{t-1,0\}}^d{\frac{a_{e,k}}{k+1}\left(\begin{array}{c}k+1\\t\end{array}\right)\Psi_{k+1-t}(N_e^Q(S))}}}\\\label{eq:alternative}
&=& \sum_e{\sum_{t=0}^{d+1}{\delta_{e,t}\Psi_t(N^P_e(S))}}.
\end{eqnarray}

By Claim \ref{claim:u-partial-potential} and the definition of the partial potential we have $\hat{c}_u(S)=\Phi_u(S)=\Phi(S)-\Phi^{{\cal N}\setminus \{u\}}(S)$. Using the alternative expression for the potentials
$\Phi(S)$ and $\Phi^{{\cal N}\setminus \{u\}}(S)$ (i.e., equality (\ref{eq:alternative})) as well as inequality (\ref{eq:use-technical}), we obtain
\begin{eqnarray*}
\hat{c}_u(S) &=& \sum_e{\sum_{k=0}^{d+1}{\delta_{e,k} \left(\Psi_k(N^{{R\cup\{u\}}}_e(S))-\Psi_k(N^{R}_e(S))\right)}}\\
&=& \sum_{e\in s_u}{\sum_{k=1}^{d+1}{\delta_{e,k} \left(\Psi_k(N^{{R\cup\{u\}}}_e(S))-\Psi_k(N^{R}_e(S))\right)}}\\
&\leq & \sum_{e\in s_u}{\sum_{k=1}^{d+1}{\delta_{e,k} \left((1+\epsilon)\Psi_k(N^{\{u\}}_e(S))+\xi_\epsilon\Psi_k(N^{{R}}_e(S))\right)}}.
\end{eqnarray*}
The second equality follows since $\Psi_0(A)=1$ for every (possibly empty) multiset of reals $A$. Using the fact again together with the fact $\Psi_k(\emptyset)=0$ for $k\geq 1$, as well as the definitions of the potentials, we obtain
\begin{eqnarray*}
\hat{c}_u(S) &\leq& (1+\epsilon)\sum_{e\in s_u}{\sum_{k=0}^{d+1}{\delta_{e,k} \left(\Psi_k(N^{\{u\}}_e(S))-\Psi_k(\emptyset)\right)}}+ \xi_\epsilon\sum_{e\in s_u}{\sum_{k=0}^{d+1}{\delta_{e,k} \left(\Psi_k(N^{R}_e(S))-\Psi_k(\emptyset)\right)}}\\
&\leq & (1+\epsilon)\sum_e{\sum_{k=0}^{d+1}{\delta_{e,k} \left(\Psi_k(N^{\{u\}}_e(S))-\Psi_k(\emptyset)\right)}}+ \xi_\epsilon\sum_{e}{\sum_{k=0}^{d+1}{\delta_{e,k} \left(\Psi_k(N^{R}_e(S))-\Psi_k(\emptyset)\right)}}\\
&=& (1+\epsilon) \left(\Phi^{{\cal N}\setminus R}(S)-\Phi^{{\cal N}\setminus (R\cup\{u\})}(S)\right)+ \xi_\epsilon\left(\Phi^{{\cal N}\setminus \{u\}}(S)-\Phi^{{\cal N}\setminus (R\cup \{u\})}(S)\right)\\
&=& (1+\epsilon) \Phi^{{\cal N}\setminus R}_u(S)+\xi_\epsilon\Phi^{{\cal N}\setminus \{u\}}_R(S)
\end{eqnarray*}
and the proof is complete.
\qed

\subsection{Proof of Lemma \ref{lem:deviation-cost-decreases-only-slightly}}
For every $i>j$ and $\epsilon>0$, we apply Lemma \ref{lem:effect} for state $(S^{i-1}_{-u},s'_u)$, player $u$, and the set $R_i$ of players that move during phase $i$ to obtain
\begin{eqnarray*}
\hat{c}_u(S^{i-1}_{-u},s'_u) & \leq & (1+\epsilon) \Phi_u^{{\cal N}\setminus R_i}(S^{i-1}_{-u},s'_u) + \xi_\epsilon\Phi_{R_i}^{{\cal N}\setminus\{u\}}(S^{i-1}_{-u},s'_u)\\
& = & (1+\epsilon) \Phi_u^{{\cal N}\setminus R_i}(S^{i}_{-u},s'_u) + \xi_\epsilon\Phi_{R_i}^{{\cal N}\setminus\{u\}}(S^{i-1})\\
& \leq & (1+\epsilon) \Phi_u(S^{i}_{-u},s'_u) + \xi_\epsilon\Phi_{R_i}(S^{i-1})\\
& = & (1+\epsilon) \hat{c}_u(S^{i}_{-u},s'_u) + \xi_\epsilon\Phi_{R_i}(S^{i-1})
\end{eqnarray*}
and, equivalently,
\begin{eqnarray*}
\hat{c}_u(S^{i}_{-u},s'_u) &\geq & \frac{1}{1+\epsilon}\hat{c}_u(S^{i-1}_{-u},s'_u)-\frac{\xi_\epsilon}{1+\epsilon}\Phi_{R_i}(S^{i-1}).
\end{eqnarray*}
The first equality in the derivation above follows by Claim \ref{claim:equal-partial-potential} since the players in ${\cal N}\setminus R_i$ use the same strategies in states $(S^{i-1}_{-u},s'_u)$ and $(S^i_{-u},s'_u)$ and since all players besides $u$ use the same strategies in states $(S^{i-1}_{-u},s'_u)$ and $S^{i-1}$. The second inequality follows by Claim \ref{claim:partial-potential-bound} and the last equality follows by Claim \ref{claim:u-partial-potential}.

We now set $\epsilon=(1+\gamma)^{1/m}-1$. This implies that $(1+\epsilon)^{-m}=(1+\gamma)^{-1}\geq 1-\gamma$. Also, by Claim \ref{claim:concave}, we get $\epsilon \geq \frac{\gamma}{m}(1+\gamma)^{1/m-1}\geq (m(1+\gamma^{-1}))^{-1}$ and, by the definition of the parameter $g$, $\xi_\epsilon= (1+m(1+\gamma^{-1})^d d^d -1 \leq \frac{g\gamma^3}{2n}$. Using the above inequality together with these observations, we obtain
\begin{eqnarray*}
\hat{c}_u(S^{m-1}_{-u},s'_u) &\geq & (1+\epsilon)^{j-m+1}\hat{c}_u(S^{j}_{-u},s'_u)-\xi_\epsilon\sum_{i=j+1}^{m-1}{(1+\epsilon)^{i-m-2}\Phi_{R_i}(S^{i-1})}\\
&\geq & (1+\epsilon)^{-m}\hat{c}_u(S^{j}_{-u},s'_u)-\xi_\epsilon\sum_{i=j+1}^{m-1}{\Phi_{R_i}(S^{i-1})}\\
&\geq &(1-\gamma)\hat{c}_u(S^{j}_{-u},s'_u)-\xi_\epsilon\sum_{i=j+1}^{m-1}{nb_i\gamma^{-1}}\\
&= & (1-\gamma)\hat{c}_u(S^{j}_{-u},s'_u)-\xi_\epsilon n\gamma^{-1}b_j\sum_{i=1}^{m-1-j}{g^{-i}}\\
&\geq & (1-\gamma) \hat{c}_u(S^{j}_{-u},s'_u)-2\xi_\epsilon n\gamma^{-1}b_jg^{-1}\\
&\geq &  (1-\gamma) \hat{c}_u(S^{j}_{-u},s'_u)-\gamma^2 b_j\\
&\geq & (1-\gamma) \hat{c}_u(S^{j}_{-u},s'_u)-\gamma \hat{c}_u(S^j)/p\\
&\geq & (1-2\gamma) \hat{c}_u(S^{j}_{-u},s'_u).
\end{eqnarray*}
The second inequality is obvious, the third inequality follows by Lemma \ref{lem:potential-bound-per-phase} and by the relation between $\epsilon$ and $\gamma$, the equality follows by the definition of $b_i$, the fourth inequality follows since $g\geq 2$ which implies that $\sum_{i\geq 1}{g^{-i}}\leq 2g^{-1}$, the fifth inequality follows by our observation about $\xi_\epsilon$ above, the sixth inequality follows since $\gamma\leq 1/p$ (this can be seen by inspecting the values of $\gamma$ and $p$ in the definition of the algorithm and the bound on $\theta_d(1+\gamma)$ provided by Lemma \ref{lem:stretch-d}) and $\hat{c}_u(S^j)$ is higher than $b_j$ when the strategy of player $u$ is irrevocably decided at the end of phase $j$, and the last inequality follows since player $u$ has no incentive to make a $p$-move at state $S^j$.
\qed

\subsection{Proof of Lemma \ref{lem:apx-bound}}
Consider the application of the algorithm to a $\Psi$-game and let $u$ be any player whose strategy is irrevocably decided at the end of phase $j$ of the algorithm. Also, let $s'_u$ be any other strategy of this player. By Lemmas \ref{lem:cost-increases-only-slightly} and \ref{lem:deviation-cost-decreases-only-slightly} and since, by the definition of the algorithm, player $u$ has no incentive to make a $p$-move at state $S^j$, we have
\begin{eqnarray*}
\frac{\hat{c}_u(S^{m-1})}{\hat{c}_u(S^{m-1}_{-u},s'_u)} &\leq & \frac{(1+2\gamma)}{(1-2\gamma)}\cdot \frac{\hat{c}_u(S^j)}{\hat{c}_u(S^j_{-u},s'_u)} \leq \frac{1+2\gamma}{1-2\gamma} \left(\frac{1}{\theta_d(1+\gamma)}-2\gamma\right)^{-1}.
\end{eqnarray*}
Hence, the right-hand side of the above inequality upper-bounds the approximation guarantee of the algorithm. For $d=1$, the parameter $\gamma$ takes values in $(0,1/10]$. Since $\gamma\in (0,1/10]$ and $\theta_1(1+\gamma)=\frac{3+\sqrt{5}}{2}+6\gamma$ (see Lemma \ref{lem:stretch-linear}), by making simple calculations, we obtain that the algorithm computes a $\hat\rho_1$-approximate equilibrium with
$$\hat\rho_1 \leq \frac{3+\sqrt{5}}{2}+110 \gamma.$$
For larger values of $d$, the algorithm uses $\gamma\in (0,\frac{1}{8\theta_d(2)}]$. Since $\theta_d(1+\gamma)$ is non-decreasing in $\gamma$, we have that $\left(\frac{1}{\theta_d(1+\gamma)}-2\gamma \right)^{-1} \leq \frac{4}{3}\theta_d(2)$. Also, we have that $\gamma < 1/34$ and hence $\frac{1+2\gamma}{1-\gamma}\leq \frac{9}{8}$. By using the value for $\theta_d(2)$ from Lemma \ref{lem:stretch-d}, we have that the algorithm computes a $\hat\rho_d$-approximate equilibrium with $\hat\rho_d \leq 3^{d+1} (d+1)^{d+1} \in d^{d+o(d)}$.
\qed

\section{The structure of the Nash dynamics of weighted congestion games with superlinear latency functions}\label{sec:modified-alg}
Algorithm 1 identifies a short sequence of best-response moves in the $\Psi$-game on input. When the degree of the $\Psi$-game is higher than $1$, the sequence may include non-improvement moves for the corresponding weighted congestion game. In this section, we present an algorithm that is applied directly to a weighted congestion game with polynomial latency functions of maximum degree $d\geq 2$. The algorithm (Algorithm 2, see the table below) is very similar to Algorithm 1; the main difference is that decisions are based on the cost of the players in the original weighted congestion game (so $\hat{c}_u$ in Algorithm 1 has been replaced by $c_u$ in Algorithm 2). In addition, the parameters $q$ and $p$ used by Algorithm 2 are higher than the ones used in Algorithm 1. The main reason is that the only available tool we have in order to guarantee convergence to an approximate equilibrium is the potential function of the corresponding $\Psi$-game. Hence, parameters $q$ and $p$ are sufficiently high so that the moves performed by Algorithm 2 are also improvement moves for the corresponding $\Psi$-game. Due to technical reasons, $\gamma$ is now restricted to smaller (but still constant) positive values. We remark that, in the description of Algorithm 2, $\BR_u$ denotes the best-response of player $u$ in the weighted congestion game.

The analysis of the algorithm will follow the same lines with the analysis of Algorithm 1. Again, the main idea in the analysis is to show that the algorithm computes an approximate equilibrium for the corresponding $\Psi$-game (with a slightly worse approximation guarantee) which is also an approximate equilibrium for the original weighted congestion game. Our main statement for Algorithm 2 is the following.

\begin{theorem}\label{thm:main-w}
For every weighted congestion game with polynomial latency functions of constant maximum degree $d\geq 2$, Algorithm 2 identifies a sequence of best-response moves from any initial state to a $\rho_d$-approximate equilibrium, where $\rho_d \in d^{O(d^2)}$. The length of the sequence is polynomial in $\gamma^{-1}$ and in the number of bits in the representation of the game.
\end{theorem}

In the following, we consider the application of the algorithm on a weighted congestion game with polynomial latency functions of degree $d$. We denote by $S^i$ the state computed by the algorithm after the execution of phase $i$ for $i=0, 1, ..., m-1$. Also, we use $R_i$ to denote the set of players that make at least one move during phase $i$. Similarly to the analysis of the algorithm for $\Psi$-games, we first aim to show that the algorithm computes a $d^{O(d^2)}$-approximate equilibrium for the corresponding $\Psi$-game. Then, the result will follow by Lemma \ref{lem:approx}.

\IncMargin{3em}
\RestyleAlgo{boxed}
\LinesNumbered
\begin{algorithm}
\SetKwData{Left}{left}\SetKwData{This}{this}\SetKwData{Up}{up}
\SetKwFunction{Union}{Union}\SetKwFunction{FindCompress}{FindCompress}
\SetKwInOut{Input}{input}\SetKwInOut{Output}{output}

\Input{A weighted congestion game ${\cal G}$ with polynomial latency functions of maximum degree $d$ with a set ${\cal N}$ of $n$ players, an arbitrary initial state $S$, and $\gamma\in (0, \left(4\cdot d! \theta_d(2(d!)^2)\right)^{-1}]$}
\Output{A state of ${\cal G}$}
   $c_{\min}\leftarrow\min_{u\in {\cal N}}{c_u(\vo_{-u},\BR_u(\vo))}$\;
   $c_{\max}\leftarrow \max_{u\in {\cal N}}{c_u(S)}$\;
   $m\leftarrow\log{\left(c_{\max}/c_{\min}\right)}$\;
   $g\leftarrow2\left(1+m(1+\gamma^{-1})\right)^{d}d^{d}n\gamma^{-3}$\;
   $q\leftarrow d! (1+\gamma)$\;
   $p \leftarrow \left(\frac{1}{d! \theta_d(d! q)} - 2\gamma\right)^{-1}$\; \label{alg-w:step2}
   \lFor{$i\leftarrow 0$ \KwTo $m$ \label{a-w}}{$b_i \leftarrow c_{\max}g^{-i}$\;} \label{alg-w:step3}
   \While{there exists a player $u\in {\cal N}$ such that $c_u(S)\in [b_1,+\infty)$ and $c_u(S_{-u},\BR_u(S))<c_u(S)/q$}
   {
   $S\leftarrow (S_{-u},\BR_u(S))$\;
   }
   $F \leftarrow \emptyset$\; \label{alg-w:step1}
   \For{phase $i\leftarrow 1$ \KwTo $m-1$ \label{main-w}}{
         \While{there exists a player $u\in {\cal N}\setminus F$ such that either $c_u(S) \in [b_{i}, +\infty)$ and $c_u(S_{-u},\BR_u(S))<c_u(S)/p$ or $c_u(S) \in [b_{i+1}, b_{i})$ and $c_u(S_{-u},\BR_u(S))<c_u(S)/q$}
         {
         $S\leftarrow (S_{-u},\BR_u(S))$\;
         }
         $F \leftarrow F \cup \{u \in {\cal N}\setminus F:c_u(S) \in [b_{i}, +\infty)\}$\;
   }
\caption{Computing approximate equilibria in weighted congestion games with polynomial latency functions.}\label{alg-w}
\end{algorithm}
\DecMargin{3em}

Again, the proof will use the same arguments as before. First, we prove the key property that the $R_i$-partial potential is small when the phase $i\geq 1$ starts. Then, we use this fact together with the parameters of the algorithm to prove that the running time is polynomial. The proof of the approximation guarantee for the corresponding $\Psi$-game follows. Again, the purpose of the third part of the proof is to show that for each player whose strategy is irrevocably decided at the end of phase $j$, neither her cost in the $\Psi$-game increases significantly nor the cost she would experience by deviating to another strategy decreases significantly after phase $j$. Hence, the approximation guarantee with respect to the $\Psi$-game in the final state computed by the algorithm is slightly higher than $p$. In our proofs, we use the terms $W$-cost and $\Psi$-cost in order to distinguish between the cost experienced by the players in the original weighted congestion game and the corresponding $\Psi$-game.

We will use the following fact that follows by Claim \ref{claim:approx}.

\begin{claim}\label{claim:modify}
Let ${\cal G}$ be a weighted congestion game with polynomial latency functions of degree $d$ and ${\cal G}'$ its corresponding $\Psi$-game. A $\rho$-move in ${\cal G}$ is a $\rho/d!$-move in ${\cal G}'$. A $\rho$-approximate equilibrium in ${\cal G}$ is a $d!\rho$-approximate equilibrium in ${\cal G}'$.
\end{claim}

\begin{proof}
Let $S$ be a state of ${\cal G}$ and consider the deviation of player $u$ to strategy $s'_u$ which is a $\rho$-move. Then, 
\begin{eqnarray*}
\hat{c}_u(S) &\geq & c_u(S) \geq  \rho c_u(S_{-u},s'_u) \geq  \frac{\rho}{d!} \hat{c}_u(S_{-u},s'_u).
\end{eqnarray*}
Now, assume that state $S$ is a $\rho$-approximate equilibrium for ${\cal G}$. For every player $u$ and every strategy $s'_u$, we have
\begin{eqnarray*}
\hat{c}_u(S) &\leq & d! c_u(S) \leq  d! \rho c_u(S_{-u},s'_u) \leq  d! \rho \hat{c}_u(S_{-u},s'_u),
\end{eqnarray*}
i.e., $S$ is a $d!\rho$-approximate equilibrium for game ${\cal G}'$.
\end{proof}

Since the parameters $q$ and $p$ used by our algorithm are strictly higher than $d!$, the above claim immediately implies that the players that move in each step actually make an improvement move in the $\Psi$-game as well.

\subsection{Proving the key property}
The key property maintained by Algorithm 2 is the following.

\begin{lemma}\label{lem:potential-bound-per-phase-w}
For every phase $i\geq 1$ of Algorithm 2, it holds that $\Phi_{R_i}(S^{i-1}) \leq \gamma^{-1} n b_{i}$.
\end{lemma}

\begin{proof}
In order to prove it, we will need Lemma \ref{lem:last-moves}. Note that the proof of Lemma \ref{lem:last-moves} works for every sequence of improvement moves by players in a set $R_i$ in a $\Psi$-game and does not depend on any particular algorithm. Since, in every phase $i$ of Algorithm 2, the players of $R_i$ do follow improvement moves in the $\Psi$-game, the proof is valid in this case as well.

Now, the argument proceeds very similarly to the proof of Lemma \ref{lem:potential-bound-per-phase}. We include the full proof here since many minor modifications are required. Again, for the sake of contradiction, we assume that $\Phi_{R_i}(S^{i-1}) > \gamma^{-1}n b_i$ and we denote by $P_i$ and $Q_i$ the set of players in $R_i$ whose last move was a $p$-move and $q$-move (in the weighted congestion game), respectively. By Claim \ref{claim:modify}, each player in $P_i$ decreases her $\Psi$-cost by at least $(p/d!-1)\hat{c}(u)$ during her last move within phase $i$. Hence, we have
\begin{eqnarray*}
\Phi_{R_{i}}(S^{i}) &\leq & \Phi_{R_{i}}(S^{i-1})- (p/d!-1)\sum_{u \in P_i} \hat{c}(u).
\end{eqnarray*}
Now, observe that each player in $Q_i$ experiences a $W$-cost of at most $b_i$ when she makes her last move within phase $i$, i.e., a $\Psi$-cost at most $d!b_i$ (by Claim \ref{claim:approx}). Using this fact and Lemma \ref{lem:last-moves}, we have
\begin{eqnarray*}
\sum_{u\in P_i}{\hat{c}(u)} &\geq & \Phi_{R_i}(S^i)-\sum_{u\in Q_i}{\hat{c}(u)} \geq \Phi_{R_i}(S^i)-d! nb_i.
\end{eqnarray*}
Using the last two inequalities and our assumption, we obtain that
\begin{eqnarray*}
\Phi_{R_i}(S^i) &\leq & \Phi_{R_i}(S^{i-1}) - (p/d!-1) \Phi_{R_i}(S^i)+(p-d!) nb_i\\
&< & (1+(p-d!)\gamma)\Phi_{R_i}(S^{i-1}) - (p/d!-1) \Phi_{R_i}(S^i)
\end{eqnarray*}
which implies that
\begin{eqnarray*}
\Phi_{R_i}(S^i) &\leq & d! \left(\frac{1}{p}+\gamma\right)\Phi_{R_i}(S^{i-1}).
\end{eqnarray*}

Now, we adapt the argument used in the proof of Lemma \ref{lem:potential-bound-per-phase} in order to reach the desired contradiction. Consider state $S^{i-1}$ and let $X_i$ and $Y_i$ be the sets of players in $R_i$ with $W$-cost at least $b_i$ and smaller than $b_i$, respectively. Notice that, by the definition of phase $i-1$ and Claim \ref{claim:modify}, $S^{i-1}$ is a $d!q$-approximate equilibrium for the players in $X_i$ (with respect to the $\Psi$-game). We construct a new $\Psi$-game of degree $d$ among the players in $\cal N$ as follows. The new game has all resources of the original game; the parameters $a_{e,k}$ for these resources are the same as in the original game. In addition, the new game has a new resource $e_u$ for each player $u\in Y_i$; the parameters for this resource are $a_{e_u,0}=d!b_i/w_u$ and $a_{e_u,k}=0$ for $k=1, ..., d$. Each player in ${\cal N} \setminus Y_i$ has the same set of strategies in the two games. The strategy set of player $u\in Y_i$ consists of the strategy $s_u$ she uses in $S^{i-1}$ as well as strategy $s'_u\cup \{e_u\}$ for each strategy $s'_u\not=s_u$ she has in the original game.

Let $\bar{S}^{i-1}$ be the state of the new game in which all players play their strategies in $S^{i-1}$. Clearly, state $\bar{S}^{i-1}$ is a $d!q$-approximate equilibrium for the players in $X_i$ (with respect to the $\Psi$-game). Also, at state $\bar{S}^{i-1}$, each player $u\in Y_i$ experiences a $\Psi$-cost equal to the $\Psi$-cost she experiences at state $S^{i-1}$ of the original game, i.e., smaller than $d!b_i$. In the new $\Psi$-game, any deviation of $u$ would include resource $e_u$ and would increase the $\Psi$-cost of player $u$ to at least $w_u a_{e_u,0}=d!b_i$. Hence, $\bar{S}^{i-1}$ is a $d!q$-approximate equilibrium for the players of $Y_i$ as well. We use $\bar\Phi$ to denote the potential of the new $\Psi$-game. Since the players use the same strategies in states $S^{i-1}$ and $\bar{S}^{i-1}$ and the parameters $a_{e,k}$ of the original resources are the same in both games, we have $\bar\Phi_{R_i}(\bar{S}^{i-1}) = \Phi_{R_i}(S^{i-1})$.

Now, let $\bar{S}^i$ be the state in which each player in ${\cal N}\setminus Y_i$ uses her strategy in $S^i$ and the strategies for the players in $Y_i$ are defined as follows. Let $u$ be a player of $Y_i$ and $s'_u$ be the strategy she uses at state $S^i$ of the original game. Her strategy in state $\bar{S}^i$ of the new $\Psi$-game is $s'_u\cup \{e_u\}$ if $s'_u\not=s_u$ and $s_u$ otherwise. Observe that, by the definition of the partial potential, we have that the partial potential $\bar\Phi_{R_i}(\bar{S}^i)$ of the new $\Psi$-game at state $\bar{S}^i$ is by at most $\sum_{u\in Y_i}{a_{e_u,0} \Psi_1(N_{e_u}(\bar{S}^i))} \leq d!nb_i$ higher than the partial potential of the original $\Psi$-game at state $S^i$ (due to the contribution of the additional resources to the potential value). Using these observations, our assumption, and the definition of parameter $p$, we have
\begin{eqnarray*}
\bar{\Phi}_{R_i}(\bar{S}^i) &\leq & \Phi_{R_i}(S^i)+d!nb_i < d!\left(\frac{1}{p}+2\gamma\right) \Phi_{R_i}(S^{i-1}) = d!\left(\frac{1}{p}+2\gamma\right) \bar{\Phi}_{R_i}(\bar{S}^{i-1}) =  \frac{1}{\theta_d(d!q)} \bar{\Phi}_{R_i}(\bar{S}^{i-1}).
\end{eqnarray*}
So, we have identified a state $\bar{S}^{i-1}$ of the new $\Psi$-game which is a $d!q$-approximate equilibrium for the players in $R_i$ and another state $\bar{S}^i$ such that the players in ${\cal N}\setminus R_i$ use the same strategies in $\bar{S}^{i-1}$ and $\bar{S}^i$ and $\bar\Phi_{R_i}(\bar{S}^{i-1}) > \theta_d(d!q) \bar\Phi_{R_i}(\bar{S}^i)$. This contradicts Lemma \ref{lem:stretch-partial} and, subsequently, it also contradicts our assumption $\Phi_{R_i}(S^{i-1})>\gamma^{-1}nb_i$. The lemma follows.
\end{proof}

\subsection{Bounding the running time}
We will now use Lemma \ref{lem:potential-bound-per-phase-w} and the properties of $\Psi$-games to prove that the algorithm terminates quickly. Again, we assume that each player can efficiently compute her best-response strategy at any state (including the pseudo-state $\vo$).

\begin{lemma}\label{lem:complexity-w}
Algorithm 2 terminates after a number of steps that is polynomial in the number of bits in the representation of the game and $\gamma^{-1}$.
\end{lemma}

\begin{proof}
At the initial state, the W-cost of each player is at most $c_{\max}$. Hence, by Claim \ref{claim:approx}, the total $\Psi$-cost of the players and, consequently (by Lemma \ref{lem:cost-vs-potential-d}), the potential of the initial state is at most $d!n\hat{c}_{\max}$. By Claim \ref{claim:modify}, each one of the players that move during phase $0$ decreases her $\Psi$-cost and, consequently (by Theorem \ref{thm:psi-potential}), the potential by at least $(q/d!-1)b_1=\gamma g^{-1} \hat{c}_{\max}$. Hence, the total number of moves in phase $0$ is at most $d!n \gamma^{-1} g$. For $i\geq 1$, we have $\Phi_{R_i}(S^i)\leq n b_i \gamma^{-1}$ (by Lemma \ref{lem:potential-bound-per-phase-w}). By Claim \ref{claim:modify}, each one of the players in $R_i$ that move during phase $i$ decreases her $\Psi$-cost and, consequently (by Claim \ref{claim:partial-potential}), the $R_i$-partial potential by at least $(q/d!-1)b_{i+1} = b_i g^{-1} \gamma$. Hence, phase $i$ completes after at most $n g \gamma^{-2}$ moves. In total, we have at most $m n g \gamma^{-2}$ moves (since $\gamma^{-1}\geq d!$). The theorem follows by observing that $g$ depends polynomially on $m$, $n$, and $\gamma^{-1}$.
\end{proof}

\subsection{Proving the approximation guarantee}
The proof of the approximation guarantee will use the following lemma (it is analogous to Lemmas \ref{lem:cost-increases-only-slightly} and \ref{lem:deviation-cost-decreases-only-slightly} in the analysis of Algorithm 1).

\begin{lemma}\label{lem:cost-changes-only-slightly-w}
Let $u$ be a player whose strategy was irrevocably decided at phase $j$ of Algorithm 2 and let $s'_u$ be any of her strategies. Then, $\hat{c}_u(S^{m-1}) \leq (1+2\gamma)\hat{c}_u(S^j)$ and $\hat{c}_u(S^{m-1}_{-u},s'_u) \geq (1-2\gamma)\hat{c}_u(S^j_{-u},s'_u)$.
\end{lemma}

\begin{proof}
The proofs of the two parts are identical to the proofs of Lemmas \ref{lem:cost-increases-only-slightly} and \ref{lem:deviation-cost-decreases-only-slightly}, respectively. All that needs to be changed is the justification of two inequalities. At the end of the proof of Lemma \ref{lem:cost-increases-only-slightly}, we used the inequality $b_j\leq \hat{c}_u(S^j)$. This holds in our case as well since the fact that the strategy of player $u$ was irrevocably decided at phase $j$ implies that $c_u(S)\geq b_j$ and, by Claim \ref{claim:approx}, we also have $\hat{c}_u(S)\geq c_u(S)$. Similarly, at the end of the proof of Lemma \ref{lem:deviation-cost-decreases-only-slightly}, we used the inequalities $\gamma b_j\leq \hat{c}_u(S^j)/p \leq \hat{c}_u(S^j_{-u},s'_u)$. What we need is essentially to show that inequality $\gamma b_j\leq \hat{c}_u(S^j_{-u},s'_u)$ holds. We have
\begin{eqnarray*}
\gamma b_j & \leq & \gamma c_u(S^j) \leq c_u(S^j)/p \leq c_u(S^j_{-u},s'_u) \leq \hat{c}_u(S^j_{-u},s'_u).
\end{eqnarray*}
The first inequality is due to the fact that the strategy of player $u$ was irrevocably decided at phase $j$, the second one follows since $\gamma\leq 1/p$, the third one follows since, at state $S^j$, player $u$ has no $p$-move in the original weighted congestion game, and the last one follows by Claim \ref{claim:approx}.
\end{proof}

We are now ready to use the last lemma in order to prove the approximation guarantee. This will complete the proof of Theorem \ref{thm:main}.
\begin{lemma}
Algorithm 2 computes a $d^{O(d^2)}$-approximate equilibrium for the weighted congestion game on input.
\end{lemma}

\begin{proof}
Consider the application of the algorithm and let $u$ be any player whose strategy is irrevocably decided at the end of phase $j$ of the algorithm. Also, let $s'_u$ be any other strategy of this player. We will show that $c_u(S^{m-1}) \leq 6 (d!)^2 \theta_d(2(d!)^2) \cdot c_u(S^{m-1}_{-u},s'_u)$; the lemma will then follow since the bound for $\theta_d(2(d!)^2)$ given by Lemma \ref{lem:stretch-d} is $d^{O(d^2)}$. We have
\begin{eqnarray*}
\frac{c_u(S^{m-1})}{c_u(S^{m-1}_{-u},s'_u)} &\leq & d!\frac{\hat{c}_u(S^{m-1})}{\hat{c}_u(S^{m-1}_{-u},s'_u)}\\
&\leq & d! \frac{(1+2\gamma)}{(1-2\gamma)}\cdot \frac{\hat{c}_u(S^j)}{\hat{c}_u(S^j_{-u},s'_u)}\\
&\leq & \frac{(1+2\gamma)}{(1-2\gamma)} \cdot d! p\\
& =& d! \frac{1+2\gamma}{1-2\gamma} \left(\frac{1}{d!\theta_d((1+\gamma)(d!)^2)}-2\gamma\right)^{-1}\\
&\leq & d! \frac{1+2\gamma}{1-2\gamma} \left(\frac{1}{d!\theta_d(2(d!)^2)}-2\gamma\right)^{-1}\\
&\leq & 2 (d!)^2 \frac{1+2\gamma}{1-2\gamma} \theta_d(2(d!)^2)\\
&\leq & 6 (d!)^2 \theta_d(2(d!)^2).
\end{eqnarray*}
The first inequality follows by Claim \ref{claim:approx}, the second one follows by Lemma \ref{lem:cost-changes-only-slightly-w}, the third one follows since, at state $S^j$, player $u$ has no $p$-move in the original weighted congestion game and, consequently (by Claim \ref{claim:modify}), no $d!p$-move in the $\Psi$-game, the equality follows by the definition of parameter $p$, the fourth inequality follows since $\theta_d$ is non-decreasing, the fifth inequality follows by the definition of parameter $\gamma$, and the last inequality follows since $\gamma\leq 1/4$.
\end{proof}


\newpage\begin{thebibliography}{99}
\bibitem{AckermannRV08}
H. Ackermann, H. R{\"o}glin, and B. V{\"o}cking.
\newblock On the impact of combinatorial structure on congestion games.
\newblock {\em Journal of the ACM}, 55(6), 2008.

\bibitem{ARV09}
H. Ackermann, H. R{\"o}glin, and B. V{\"o}cking.
\newblock Pure Nash equilibria in player-specific and weighted congestion games.
\newblock {\em Theoretical Computer Science}, 410(17): 1552--1563, 2009.

\bibitem{AC09}
E. Anshelevich and B. Caskurlu. Exact and approximate equilibria for optimal group network formation. In {\em Proceedings of the 17th Annual Symposium on Algorithms (ESA)}, LNCS 5757, Springer, pages 239--250, 2009.

\bibitem{AwerbuchAEMS08}
B. Awerbuch, Y. Azar, A. Epstein, V.~S. Mirrokni, and A.
  Skopalik.
\newblock Fast convergence to nearly optimal solutions in potential games.
\newblock In {\em Proceedings of the 9th ACM Conference on Electronic Commerce (EC)}, ACM, pages 264--273,
  2008.

\bibitem{BhalgatCK10}
A. Bhalgat, T. Chakraborty, and S. Khanna.
\newblock Approximating pure Nash equilibrium in cut, party affiliation, and
  satisfiability games.
\newblock In {\em Proceedings of the 11th ACM Conference on Electronic Commerce (EC)}, ACM, pages 73--82.
  2010.

\bibitem{C09}
I. Caragiannis. Efficient coordination mechanisms for unrelated machine scheduling. In {\em Proceedings of the 20th Annual ACM-SIAM Symposium on Discrete Algorithms (SODA)}, pp. 815--824, 2009. Full version available at {\tt arxiv.org}.

\bibitem{CFGS11}
I. Caragiannis, A. Fanelli, N. Gravin, and A. Skopalik. Efficient computation of approximate pure Nash equilibria in congestion games. In {\em Proceedings of the 52nd Annual IEEE Symposium on Foundations of Computer Science (FOCS)}, pp. 532-541, 2011. Full version available at {\tt arxiv.org}.

\bibitem{CH10}
J. Cardinal and M. Hoefer. Non-cooperative facility location and covering games. {\em Theoretical Computer Science}, 411(16-18): 1855--1876, 2010.

\bibitem{CKM+08}
M. Charikar, H. J. Karloff, C. Mathieu, J. Naor, and M. E. Saks. Online multicast with egalitarian cost sharing. In {\em Proceedings of the 20th Annual ACM Symposium on Parallelism in Algorithms and Architectures (SPAA)}, ACM, pages 70--76, 2008.

\bibitem{CCL+07}
C. Chekuri, J. Chuzhoy, L. Lewin-Eytan, J. Naor, and A. Orda. Non-cooperative multicast and facility location games. {\em IEEE Journal on Selected Areas in Communications}, 25(6): 1193--1206, 2007.

\bibitem{ChienS07}
S. Chien and A. Sinclair.
\newblock Convergence to approximate Nash equilibria in congestion games.
\newblock {\em Games and Economic Behavior}, 71(2): 315--327, 2011.

\bibitem{CGC04}
N. Christin, J. Grossklags, and J. Chuang. Near rationality and competitive equilibria in networked systems. In {\em Proceedings of the ACM SIGCOMM Workshop on Practice and Theory of Incentives in Networked Systems (PINS)}, ACM, pages 213--219, 2004.

\bibitem{CMS06}
G. Christodoulou, V. S. Mirrokni, and A. Sidiropoulos. Convergence and approximation in potential games. In {\em Proceedings of the 23rd Annual Symposium on Theoretical Aspects of Computer Science (STACS)}, LNCS 3884, Springer, pages 349--360, 2006.

\bibitem{DP07}
C. Daskalakis and C. H. Papadimitriou. Computing equilibria in anonymous games. In {\em Proceedings of the 48th Annual IEEE Symposium on Foundations of Computer Science (FOCS)}, IEEE, pages 83--93, 2007.

\bibitem{DS08}
J. Dunkel and A. S. Schulz. On the complexity of pure-strategy Nash equilibria in congestion and local-effect games. {\em Mathematics of Operations Research}, 33(4): 851--868, 2008.

\bibitem{EKM07}
E. Even-Dar, A. Kesselman, and Y. Mansour. Convergence time to Nash equilibrium in load balancing. {\em ACM Transactions on Algorithms}, 3(3), 2007.

\bibitem{FKS05}
D. Fotakis, S. Kontogiannis, and P. G. Spirakis. Selfish unsplittable flows. {\em Theoretical Computer Science}, 340(3), pp. 514--538, 2005.

%\bibitem{HLP52}
%G. Hardy, J. E. Littlewood, and G. P\'olya. {\em Inequalities}. 2nd edition, Cambridge University Press, 1952.

\bibitem{FabrikantPT04}
A. Fabrikant, C.~H. Papadimitriou, and K. Talwar.
\newblock The complexity of pure nash equilibria.
\newblock In {\em Proceedings of the 36th Annual ACM Symposium on Theory of Computing (STOC)}, ACM, pages 604--612, 2004.

%\bibitem{FanelliFM08}
%A. Fanelli, M. Flammini, and L. Moscardelli.
%\newblock The speed of convergence in congestion games under best-response
%  dynamics.
%\newblock In {\em Proceedings of the 35th International Colloquium on Automata, Languages and Programming (ICALP)}, Part 1, LNCS 5125, Springer, pages 796--807, 2008.

\bibitem{FM09}
A. Fanelli and L. Moscardelli. On best response dynamics in weighted congestion games with polynomial delays. In {\em Proceedings of the 5th International Workshop on Internet and Network Economics (WINE)}, LNCS 5929, Springer, pages 55--66, 2009.

\bibitem{GMV05}
M. X. Goemans, V. S. Mirrokni, and A. Vetta. Sink equilibria and convergence. In {\em Proceedings of the 46th Annual IEEE Symposium on Foundations of Computer Science (FOCS)}, pp. 142-154, 2005.

\bibitem{HK10}
T. Harks and M. Klimm. On the existence of pure Nash equilibria in weighted congestion games.
\newblock In {\em Proceedings of the 37th International Colloquium on Automata, Languages and Programming (ICALP)}, Part 1, LNCS 6198, Springer, pages 79--89, 2010.

\bibitem{Johnson88}
D.~S. Johnson, C.~H. Papadimitriou, and M. Yannakakis.
\newblock How easy is local search?
\newblock {\em Journal of Computer and System Sciences}, 37: 79--100, 1988.

\bibitem{KR11}
K. Kollias and T. Roughgarden. Restoring pure equilibria to weighted congestion games. In {\em Proceedings of the 38th International Colloquium on Automata, Languages and Programming (ICALP)}, Part II, LNCS 6756, Springer, pages 539--551, 2011.

%\bibitem{LO01}
%L. Libman and A. Orda. Atomic resource sharing in noncooperative networks. {\em Telecommunication Systems}, 17(4): 385-409,  2001.

\bibitem{MV04}
V. S. Mirrokni and A. Vetta. Convergence issues in competitive games.
In {\em Proceedings of the 7th International Workshop on Approximation Algorithms for Combinatorial Optimization Problems (APPROX)},
LNCS 3122, Springer, pages 183--192, 2004.

%\bibitem{Monderer96}
%D. Monderer and L.~S. Shapley.
%\newblock Potential games.
%\newblock {\em Games and Economic Behavior}, 14(1):124--143, 1996.

\bibitem{NT09}
T. Nguyen and E. Tardos. Approximate pure Nash equilibria via Lov\'asz local lemma. In {\em Proceedings of the 5th International Workshop on Internet and Network Economics (WINE)}, LNCS 5929, Springer, pages 160--171, 2009.

\bibitem{PS06}
P. N. Panagopoulou and P. G. Spirakis. Algorithms for pure Nash equilibria in weighted congestion games. {\em ACM Journal of Experimental Algorithmics}, 11, 2006.

\bibitem{R73}
R.~W. Rosenthal.
\newblock A class of games possessing pure-strategy Nash equilibria.
\newblock {\em International Journal of Game Theory}, 2:65--67, 1973.

\bibitem{SkopalikV08}
A. Skopalik and B. V{\"o}cking.
\newblock Inapproximability of pure Nash equilibria.
\newblock In {\em Proceedings of the 41st Annual ACM Symposium on Theory of Computing (STOC)}, ACM, pages 355--364, 2008.

\end{thebibliography}
\end{document}